\documentclass[12pt]{article}
\usepackage[psamsfonts]{amssymb}
\usepackage{cmmib57}
\usepackage{amsmath,amssymb,amscd}
\newcommand{\bPf}{\par\vspace*{-4pt}\indent{\sc Proof.}\enskip}
\newcommand{\ePf}{\medskip}
\def\QED{\hskip0.1em\hfill\null\ \null\nobreak\hfill\kern3pt\vbox{\hrule\hbox
   {\vrule\kern1pt\vbox{\kern1.7pt\hbox{$\scriptscriptstyle{QED}$}
    \kern0.2pt}\kern1pt\vrule}\hrule}}

\def\END{\hskip0.1em\hfill\null\ \null\nobreak\hfill\kern3pt\vbox{\hrule\hbox
   {\vrule\kern1pt\vbox{\kern1.7pt\hbox{$\,\,\,\vspace{5pt}$}
    \kern0.2pt}\kern1pt\vrule}\hrule}}
\newtheorem{theorem}{Theorem}
\newtheorem{lemma}{Lemma}
\newtheorem{corollary}{Corollary}
\newtheorem{proposition}{Proposition}
\newtheorem{remark}{Remark}
\newtheorem{definition}{Definition}
\newtheorem{example}{Example}

\newcommand{\bCd}{\bEq\begin{CD}}
\newcommand{\eCd}{\end{CD}\eEq}
\newcommand{\bcd}{\beq\begin{CD}}
\newcommand{\ecd}{\end{CD}\eeq}
\newcommand{\ben}{\begin{enumerate}}
\newcommand{\een}{\end{enumerate}}
\newcommand{\bEq}{\begin{eqnarray}}
\newcommand{\eEq}{\end{eqnarray}}
\newcommand{\beq}{\begin{eqnarray*}}
\newcommand{\eeq}{\end{eqnarray*}}
\newcommand{\bDf}{\begin{definition}\em}
\newcommand{\eDf}{\end{definition}}
\newcommand{\bLm}{\begin{lemma}}
\newcommand{\eLm}{\end{lemma}}
\newcommand{\bPr}{\begin{proposition}}
\newcommand{\ePr}{\end{proposition}}
\newcommand{\bTh}{\begin{theorem}}
\newcommand{\eTh}{\end{theorem}}
\newcommand{\bCr}{\begin{corollary}}
\newcommand{\eCr}{\end{corollary}}
\newcommand{\bRm}{\begin{remark}\em}
\newcommand{\eRm}{\end{remark}}
\newcommand{\bEx}{\begin{example}\em}
\newcommand{\eEx}{\end{example}}

\newcommand{\Z}{\mathbb{Z}}

\newcommand{\R}{I\!\!R}

\newcommand{\der}{\partial}

\newcommand{\bOme}{\boldsymbol{\Ome}}

\newcommand{\cP}{\mathcal{P}}

\newcommand{\cS}{\mathcal{S}}

\newcommand{\bx}{\boldsymbol{x}}

\newcommand{\bB}{\boldsymbol{B}}
\newcommand{\bC}{\boldsymbol{C}}

\newcommand{\bG}{\boldsymbol{G}}

\newcommand{\bM}{\boldsymbol{M}}

\newcommand{\bP}{\boldsymbol{P}}

\newcommand{\bR}{\boldsymbol{R}}

\newcommand{\bU}{\boldsymbol{U}}

\newcommand{\bX}{\boldsymbol{X}}
\newcommand{\bY}{\boldsymbol{Y}}

\newcommand{\sub}{\subset}

\newcommand{\ten}{\!\otimes\!}

\newcommand{\Ome}{\Omega}

\newcommand{\Var}{{\mathcal{V}}}

\title{\large{\bf Variational cohomology and topological solitons in Yang--Mills--Chern--Simons theories}}
\author{{\normalsize Ekkehart Winterroth}
\\
{\footnotesize Department of Mathematics, University of Torino, Italy}
\\ 
{\footnotesize via C. Alberto 10, 10123 Torino, Italy}
\\
{\footnotesize  and  Lepage Research Institute, 17 November 1, 081 16 Pre\v sov,
Slovak Republic}
\\
{\footnotesize e--mail: {\sc  ekkehart.winterroth@unito.it}}
}
\date{}
\begin{document}
\maketitle

\begin{abstract}
In cohomological formulations of the calculus of variations obstructions to the existence of (global) solutions of the Euler--Lagrange equations can arise in principle. It seems, however, quite common to assume that such obstructions always vanish, at least in the cases of interest in theoretical physics. This is not so: for Yang--Mills--Chern--Simons theories  on compact manifolds in odd dimensions $\geq 5$ we find a non trivial obstruction which leads to a quite strong non existence theorem for topological solitons/instantons. The consequences of this result for the Yang--Mills--Chern--Simons theories of holographic QCD (on $\R^5$) are discussed.
\end{abstract}


\noindent {\bf 2020 MSC primary}: 53C07; 53Z05; 57R20;

\noindent {\bf 2020 MSC secondary}: 81T13; 81T20; 

\section{Introduction}

By ``variational cohomology'' we refer, somewhat loosely, to any formulation of the calculus of variations in terms of a (co-)chain complex or double complex derived from the de Rham complex of a fibre bundle $\bY \overset{\pi}{\rightarrow} \bX$ such that the objects and operations of the calculus of variations (Lagrangians, equations, conservation laws and conserved quantities, taking the variation, etc.) are expressed in terms of its elements and (co-)boundary operators, while the complex still calculates de Rham cohomology. Variational cohomology is part of the general approach to mathematical/theoretical phyiscs for which the expression ``Cohomological Physics'' has been coined (see \cite{Sta88}).

Since de Rham cohomology detects obstructions to glue local solutions of differential equations together to global solutions, it is natural to ask if in the cohomological formulations of the calculus of variations de Rham cohomology classes appear as cohomological obstructions to the existence of solutions of Euler--Lagrange equations.
Regarding this, the common conviction seems to be that for variational problems relevant in theoretical physics such obstructions vanish always. This is not the case.
In \cite{PaWi16} (see also section \ref{Obs} below) we introduced a family of cohomological obstructions  $[\Xi \rfloor\eta_\lambda]$ and showed that the well known fact that $3$--dimensional Chern--Simons theories admit solutions only on flat bundles can be expressed by means of these.

Chern--Simons theories are Lagrangian gauge field theories for principal connections derived from the Chern--Simons secondary characteristic class corresponding to some Chern--Weil characteristic class. They have two principal applications: in quantum gravity to overcome certain problems with General Relativity (and to facilitate the unification of {\em all} fundamental fields), see e.g. \cite{BaVeCa20,Zan12}, and in the so called Yang--Mill--Chern--Simons theories, see below.
They are local variational problems, i.e. they consist  of a collection of local Lagrangians, e.g. one for each coordinate patch, which, nevertheless, have unique global Euler--Lagrange equations; in other words, they are related like a closed form and a set of local potentials (see e.g. \cite{FePaWi10}).

Here, we continue the investigation of non trivial obstructions arising in variational cohomology for Chern--Simons theories in higher dimensions. More specifically, we analyze Lagrangian gauge field theories derived from the $p+1$--component of the Chern character for principal connections on a $U(n)$--principal bundle over a $2p+1$-dimensional manifold, $n \geq p$ (for the case of theories derived from the $p+1$th Chern class see \cite{PaWi22}). As in \cite{PaWi22}, we find a non trivial obstruction which vanishes if and only if the $p$--component of the Chern character vanishes (as a cohomology class). What is exactly what one should expect in view of the $3-$dimensional case mentioned above and treated in \cite{PaWi16}. 

This obstruction has its natural field of application in Yang--Mill--Chern--Simons theories. YMCS theories are Lagrangian gauge field theories the Lagrangian of which is a sum of the YM-- and CS--Lagrangians; this construction seems to be mainly motivated by the r\^ole of the Chern--Simons form in anomaly physics, see e.g. \cite{AGG85,BaZu84,Zum84}. We derive a a quite strong non existence theorem for solitons/instantons in theories on compact mnaifolds. It raises doubts about the mathematical consistency of their use in theories of nuclear matter in five dimensional (holographic) QCD,  as in e.g. \cite{BaBo17,BRW10,KMS12,KMS15,PoWu09,RSZ10}, even on $\R^5$: The obstruction cannot vanish for $SU(n)$--theories and in the case of $U(n)$--theories mathematical inconsistencies in the asymptotic behaviour of the $U(1)$--component may arise. However controlling these possibiliteis in specific theories is quite complicated, we illustrate this in case of the Sakai--Sugimoto model, widely considered the most important example of YMCS theory for holographic QCD.

The paper is organized as follows. In section \ref{VarCo} we set up the calculus of variations on fibre bundles, the particular cohomological formulation of the calculus of variations we will use and introduce our obstructions; we also introduce the concept of local variational problem. 

Since variational cohomology requires variational problems to be formulated in terms of sections of fibre bundles, in section \ref{ConBun} we construct the (so called) bundle of connections of a principal bundle.

In section \ref{CSBun} we set up the formulation of Chern--Simons theories on the bundle of connections via Chern--Weil theory and Chern--Simons secondary characteristic classes. 

Section \ref{Char} deals with the $p+1$--component of the Chern character for principal connections on a $U(n)$--principal bundle over a $2p+1$-dimensional manifold, $n \geq p$. First, we identify a certain cohomological obstruction in Chern--Simons theories and then we analyze its effects in Yang--Mill--Chern--Simons theories. Here, the power of variational cohomology is on display: it extracts from the Yang--Mills Euler--Lagrange form a specific counterpart to the obstruction. For the existence of solutions it is then necessary that the counterpart and the corresponding obstruction annihilate each other. This leads to a rather strong non existence theorem for solitons/instantons in Yang--Mill--Chern--Simons theories on compact manifolds. We conclude the section by discussing the consequences of this non existence theorem for the Yang--Mill--Chern--Simons theories in five dimensional (holographic) QCD.
 
\section{Variational Cohomology} \label{VarCo}
We will now outline the construction and the characteristics of Krupka's variational sequence on finite order jet bundles following \cite{Kru15} (see also \cite{Kru90}) the particular variational cohomology theory which we will use. The core of the construction consists in taking the quotient of the de Rham complex of some jet bundle $J^{q} \bY$  of a fibre bundle $\bY \rightarrow \bX$ by a suitable subcomplex with the idea of eliminating the contact forms, i.e. the forms which vanish along sections, in the right way: not {\em all} contact forms, since for $p>$ dim$\bX$, $\bOme^p (J^k \bY)$ consists, of course, entirely of contact forms for dimensional reasons. The particular feature of Krupka's variational sequence is that the resulting  chain complex comes equipped by construction with an explicit representation as a complex of differential forms on some $J^{k} \bY, k > q$. 

We will conclude this section explaining the cohomological obstructions introduced in \cite{PaWi16}.

\subsection{Jet bundles}
For a fibre bundle $\bY \overset{\pi}{\rightarrow} \bX$ the {\bf $k$th order jet manifold} is defined as the set
$
J^k \bY := \cup_ {x \in \bX} \, j_x^k \sigma
$
. $j^k_x \sigma$ are the equivalence classes of germs of sections $\sigma: \bX \rightarrow \bY$ of $\pi$ at $x \in \bX$, where two germs are equivalent if and only if their Taylor expansions at $x$ coincide up to the $k$-th order (note that $J^0 \bY = \bY$).

$J^k \bY$ has a natural manifold structure (see \cite{Saund}). The derivatives up to order $k$ of $\sigma$ at $x$ define coordinate charts: 
\bEq \label{JetCor}
( x^\mu, y^i, y^i_\alpha, \dots, y^i_{\alpha_1 \dots \alpha_k} ) := \left( x^\mu, \sigma^i(x), \frac{\der \sigma^i(x)}{\der x_{\alpha}}, \dots, \frac{\der^k \sigma^i(x)}{\der x_{\alpha_1} \dots \der x_{\alpha_k}} \right)
\eEq
where $( x^\mu, y^i)$ are fibered coordinates on $\bY$ and it is to be understood that the indices are ordered: $\alpha_1 \leq \dots \leq \alpha_k$, while in a coordinate expression like $y^i_{\alpha_1 \dots \alpha_{k} \mu }$ it is to be understood that the index $\mu$ is placed on its proper place in the ordering of the indices. 

$J^k \bY$, the {\bf $k$th order jet bundle}, is a fibre bundle over $\bY$ and we denote its projection by $\pi^k: J^k \bY \rightarrow \bY$
Furthermore, we have the following topological (but not linear) isomorphism
$$
J^k \bY = \pi^*\left(\Sigma_{i=1}^{k} \cS^i \left(T^* \bX \right)\right)\otimes V\bY
$$
Here, $\cS^i \left(T^* \bX \right)$ is the $i$th symmetric power of $T^* \bX$ and $V\bY$ is the vertical bundle with respect to $\pi$.

The natural projections $J^k \bY \overset{\pi^k_{k-1}}{\rightarrow} J^{k-1} \bY$ are affine bundles.
For every section $\sigma: \bX \rightarrow \bY$, local or global, there is by construction a section $j^k \sigma: \bX \rightarrow J^k \bY$, the $k$th order prolongation of $\sigma$:
\bEq \label{JetCor2}
j^k \sigma := \left( x^\mu, \sigma^i(x), \frac{\der \sigma^i(x)}{\der x_{\alpha}}, \dots, \frac{\der^k \sigma^i(x)}{\der x_{\alpha_1} \dots \der x_{\alpha_k}} \right)
\eEq
Note that  $\bY$ is a retract of $J^k \bY$ and, thus, their cohomologies are isomorphic; furthermore, the images in cohomology of $\pi^k$ and any section of $J^k \bY \rightarrow \bY$ introduce a canonical isomorphism.
\subsubsection{The contact structure} \label{candec}
The differential at some point $x \in \bX$ of the $k$th order prolongation of some section $\sigma$
$$
d(j^{r}  \sigma(x)): T_x \bX \rightarrow ((\pi^{r+1} _r)^* T (J^r \bY )_{j^{r}  \sigma(x)})
$$
has the coordinate expression 
$$
d(j^{r}  \sigma(x))(\xi) = \sum\limits_\mu  \,\,\xi^\mu (x) \left( \frac{\der}{\der x^\mu} + \sum\limits_{k=0}^r \, \sum\limits_{i} \, \sum\limits_{\alpha_1 \leq \dots \leq \alpha_k}\, \frac{\der^{k+1} \sigma^i(x)}{\der x_\mu \der x_{\alpha_1} \dots \der x_{\alpha_k}} \, \frac{ \der} { \der y^i_{\alpha_1 \dots \alpha_k}} \right)
$$
Comparing this to the expressions \ref{JetCor} and \ref{JetCor2} above, we see that the image of $T_x \bX $ can be naturally seen as a subspace of the pullback of the tangent bundle of $J^r \bY$ to $J^{r+1} \bY$ at $j^{r+1}  \sigma(x)$. 
Collecting all these subspaces into a subbundle we obtain the {\bf canonical horizontal complement} to the pullback of the $\pi^r$-vertical subbundle 
\bEq \label{candec1}
(\pi^{r+1}_r)^* T(J^r \bY) = (\pi^{r+1}_r)^* V(J^r \bY) \oplus H (\pi^{r+1}_r)
\eEq
We have the canonical projection $h: T(J^r \bY) \mapsto H (\pi^{r+1}_r)$, the {\bf horizontalization}. For a vector field on $J^r \bY$ with the local coordinate expression
$$
\xi = \sum\limits_\mu  \,\,\xi^\mu \cdot \frac{\der}{\der x^\mu} + \sum\limits_{k=0}^r \, \sum\limits_{i} \, \sum\limits_{\alpha_1 \leq \dots \leq \alpha_k}\, \Xi_{\alpha_1 \dots \alpha_{k}}^i \cdot \frac{ \der}{ \der y^i_{\alpha_1 \dots \alpha_k}}
$$
where $\xi^\mu$ and $\Xi_{\alpha_1 \dots \alpha_{k}}^i$ are functions on the respective coordinate patch of $J^r \bY$,  it has the local coordinate expression 
$$
h(\xi)= \sum\limits_\mu  \,\,\xi^\mu \cdot \left( \frac{\der}{\der x^\mu} + \sum\limits_{k=0}^r \, \sum\limits_{i} \, \sum\limits_{\alpha_1 \leq \dots \leq \alpha_k}\, y^i_{\alpha_1 \dots \alpha_{k} \mu} \cdot \frac{ \der} { \der y^i_{\alpha_1 \dots \alpha_k}} \right)
$$
Likewise we get the complementary canonical projection $p_1: T(J^r \bY) \mapsto (\pi^{r+1}_r)^* V(J^r \bY)$ with the local coordinate expression
$$
p_1(\xi) = \sum\limits_{k=0}^r \, \sum\limits_{i} \, \sum\limits_{\alpha_1 \leq \dots \leq \alpha_k}\, \left(\Xi_{\alpha_1 \dots \alpha_{k}}^i - \sum\limits_{\mu} \, y^i_{\alpha_1 \dots \alpha_{k} \mu} \cdot \xi^\mu (x) \right) \cdot \frac{ \der}{ \der y^i_{\alpha_1 \dots \alpha_k}}
$$
For differential $1$-forms $\gamma$ the horizontalization 
$h$
is defined by 
\beq
h(\gamma) (\pi^{r+1}_{r} \circ \xi) = (\pi^{r+1}_{r})^*(\gamma)(h(\xi))
\eeq
We have  
$$
h(dx^\mu) = dx^\mu \qquad h(dy^i_{\alpha_1 \dots \alpha_k})=\sum\limits_\mu \, y^i_{\alpha_1 \dots \alpha_k \mu}dx^\mu
$$
The $\theta$ for which $h(\theta) = 0$ are called {\bf contact $1$-forms} of order $r$. They are characterized by the fact that their pullback with the $r$th-order prolongation of a section $j^r\sigma: \bX \rightarrow J^r \bY$ vanishes, i.e. being contact does not depend on the choice of a coordinate chart.
The complementary projection $p_1$ of the horizontalization 
is then defined by 
\beq
p_1(\gamma) (\pi^{r+1}_{r} \circ \xi) = (\pi^{r+1}_{r})^*(\gamma)(p_1(\xi))
\eeq
Locally, we have
$$
p_1(dx^\mu) = 0 \qquad 
p_1(dy^i_{\alpha_1 \dots \alpha_k}) = dy^i_{\alpha_1 \dots \alpha_k} - \sum\limits_\mu \, y^i_{\alpha_1 \dots \alpha_k \mu}dx^\mu \,\, =: \theta^i_{\alpha_1 \dots \alpha_k}
$$ 
with $0 \leq k \leq r$. The case $k = 0$ is understood to be
$$
p_1(dy^i) = dy^i - \sum\limits_\mu \, y^i_{ \mu}dx^\mu \,\, =: \theta^i
$$
Let $1 \leq \mu \leq dim \bX$, \, $0 \leq k \leq r-1$ and $1 \leq i \leq dim \bY-dim \bX \,$, the collection
$$
dx^\mu, \,\, \theta^i_{\alpha_1 \dots \alpha_k}, \,\, dy^i_{\alpha_1 \dots \alpha_r}
$$
is a local basis of $\bOme^1 (J^r \bY)$. Only in $(\pi^{r+1}_r)^*\bOme^1 (J^r \bY)$) we have
\beq
(\pi^{r+1}_r)^*dy^i_{\alpha_1 \dots \alpha_r} = \sum\limits\mu \, y^i_{\alpha_1 \dots \alpha_r \mu}dx^\mu + \theta^i_{\alpha_1 \dots \alpha_r}
\eeq
 Therefore, $(\pi^{r+1}_r)^* \bOme^1 (J^r \bY)$ has a basis
$$
dx^\mu, \,\, \theta^i_{\alpha_1 \dots \alpha_k} 
$$
where instead $ \, 0 \leq k \leq r $. 
For differential $q$-forms $\gamma$ in $\boldsymbol{\Omega}^k (J^{r} \bY)$ we have then
$$
(\pi^{r+1}_{r})^*(\gamma) (\pi^{r+1}_{r} \circ \xi_{1}, \dots , \pi^{r+1}_{r} \circ \xi_{q}) =  (\pi^{r+1}_{r})^*(\gamma)(h(\xi_{1}) + p_1(\xi_{1}), \dots , h(\xi_{k}) + p_1(\xi_{q}))
$$
Therefore, there is a canonical decomposition 
$$
(\pi^{r+1}_{r})^*(\gamma) = \gamma_0 + \dots + \gamma_q
$$
 $\gamma_0 = h(\gamma)$ and for $1 \leq k \leq q$
$$
\gamma_k = \sum\limits_{i_1 \leq \dots\leq i_k, \, i_{k+1} \leq \dots \leq i_q}\, |I|\cdot(\pi^{r+1}_{r})^*(\gamma)(p_1(\xi_{i_1}), \dots , p_1(\xi_{i_k}), h(\xi_{k+1}), \dots ,h(\xi_{q}))
$$
here $|I|$ is the sign of the permutation between $(1, \dots, q)$ and $i_1, \dots, i_k, i_{k+1},\dots, i_q$. $\gamma_k$ is called the $k$-contact component. Locally it is a linear combination of forms of type
$$
\delta \, \wedge \, \theta^i_{\alpha_1 \dots \alpha_j} \wedge \dots \wedge d\theta^i_{\alpha_1 \dots \alpha_{r-1}}
$$
where $\delta$ is a degree $q-k$ horizontal form and the exterior product of the $\theta^i_{\alpha_1 \dots \alpha_k}$  and the $d\theta^i_{\alpha_1 \dots \alpha_{r-1}}$ is of degree $k$
$$
p_k(\gamma):= \gamma_k
$$
 is the projection  of $\gamma$ on its $k$-contact component. The fact that for $q \geq n+1$ all $q$-forms are contact is reflected and refined by the following propety of the canonical decomposition: let $q = n+k$ and $k \geq 1$, we have then
$$
\gamma_0 = \dots = \gamma_{k-1} = 0 \qquad (\pi^{r+1}_{r})^*(\gamma) = \gamma_k + \dots + \gamma_q
$$
A differential form is {\bf $k$-contact} if and only if $\gamma_0 = \dots = \gamma_{k-1} = 0$ and $\gamma_k \neq 0$.
The horizontalization allows also to define the "prolongation" of the partial derivatives of $\bX$, the {\bf total} or {\bf formal derivative} $d_\mu$ by
$$
h(df) = \sum\limits_\mu  \,\, d_\mu f \cdot dx^\mu
$$
explicitly, we have  
\bEq \label{totder}
d_\mu = h(\frac{\der}{\der x^\mu}) =  \frac{\der}{\der x^\mu} + \sum\limits_{k=0}^r \, \sum\limits_{i} \, \sum\limits_{\alpha_1 \leq \dots \leq \alpha_k}\, y^i_{\alpha_1 \dots \alpha_{k} \mu} \, \frac{ \der} {\der y^i_{\alpha_1 \dots \alpha_k}}
\eEq
\subsection{Variational problems on fibre bundles} \label{VarBun}
Before turning to the variational sequence we need to sketch out the basics of the variational calculus on fibre bundles. First of all, from now on we set $dim \, \bX = n$ and  $dim \, \bY = n + m$.
A $r$-th order Lagrangian is a horizontal $n$-form $\lambda$ on $J^r \bY$, in coordinates
$$
\lambda  = \mathcal{L}(x^\mu, y^i, y^i_\alpha, \dots, y^i_{\alpha_1 \dots \alpha_r}) dx_1 \wedge \dots \wedge dx_n \,.
$$
 
The corresponding variational problem consists in finding the critical sections $\sigma$ of $\bY \overset{\pi}{\rightarrow} \bX$, i.e. the sections for which the integral 
\bEq \label{VarBun1}
\int_{\bX} j^r \sigma^* \mathcal{L}
\eEq 
over $\bX$ itself or the integrals
\bEq \label{VarBun2}
\int_{\bB} j^r \sigma^* \mathcal{L}
\eEq
over all $n$-dimensional compact submanifolds with boundary  $\bB$ of $\bX$ take their extremal values, i.e. in the calculus of variations on fibre bundles the variation of the functional is induced by the variation of the section of the bundle. 
On $J^{2r+1} \bY$ we have then the $n+1$-form $E_{\lambda}$, the {\bf Euler-Lagrange form},
 locally 
$$
E_{\lambda}  = \sum\limits_{i} \, E( \mathcal{L})_i \, \theta^i \wedge dx_1 \wedge \dots \wedge dx_n
$$ 
with 
$$
E( \mathcal{L})_i  = \frac{\der  \mathcal{L}}{\der y_i} + \sum\limits_{k=0}^{r+1} \,\, (-1)^k \, d_{\alpha_1} \dots d_{\alpha_k} \, \frac{\der  \mathcal{L}}{\der y^i_{\alpha_1 \dots \alpha_k} }
$$
(here  $\theta^i = dy^i - \sum\limits_\mu y^i_{\mu}dx^\mu$, see above). 
The critical sections  satisfy then the {\bf Euler-Lagrange equations}:
\beq 
 E_{\lambda}\circ j^{2r+1} \sigma  = 0
\eeq
This is, of course, {\em not} the pullback $j^{2r+1} \sigma^* \left(E_{\lambda}\right)$ of the differential $n+1$-form $E_{\lambda}$, which would be zero for any section $\sigma$ for dimensional reasons, but the restriction $E_{\lambda}|_{j^{2r+1} \sigma}$. Analogously, one defines {\bf Helmholtz conditions}, the {\bf first variation formula}, etc. for the calculus of variations on fibre bundles. 
A {\bf local variational problem} of order $r$ on $\bY$ consists in an open covering  $\mathcal{U}$ of $\bY$ and a $r$th-order Lagrangian $\lambda_i$ on each $U_i \in \mathcal{U}$ such that there is a globally well defined $n+1$-form $\eta_{\lambda}$ on $J^{2r+1} \bY$ with 
\beq 
\eta_{\lambda}|_{J^{2r+1} U_i} = E_{\lambda_i}
\eeq
\subsection{The Variational Sequence}
The observation that contact forms do not contribute to the action integral, see (\ref{VarBun1}) and (\ref{VarBun2}) above, is the starting point for the construction of any cohomological formulation of the calculus of variations.
In the case of the variational sequence the construction  consists in taking the quotient of the de Rham complex $(\bOme^* (J^{r-1} \bY), d)$ of $J^{r-1} \bY$ by a certain cohomologically trivial subcomplex $(\boldsymbol{\Theta}^{*}_{r-1}, d)$ of contact forms and finding a representation of the quotient 
$$
(\bOme^* (J^{r-1} \bY) / \boldsymbol{\Theta}^{*}_{r-1}, \, \widehat{d}\, ) \mapsto (\boldsymbol{\Var}^{*}_{r}, E_{i})
$$
as a complex of differential forms $(\boldsymbol{\Var}^{*}_{r}, E_{i})$ on higher order jet bundles of $\bY$. Both the quotient and its representation will still calculate the de Rham cohomology of $J^{r-1} \bY$ (canonically isomorphic to that of $\bY$). The $\lambda \in \boldsymbol{\Var}^{n}_{r}$ are then the Lagrangians and $E_{n} (\lambda) \in \boldsymbol{\Var}^{n+1}_{r} \sub \bOme^{n+1} (J^{2r-1} \bY)$ are the Euler-Lagrange forms. We begin with $J^{r-1} \bY$ instead of $J^{r} \bY$, so we cover only that type of $r$th-order Lagrangians we are actually interested in (see the remarks at the end).

The construction is sheaf theoretic in nature and not very enlightening: many steps involve rather cumbersome calculations and are intelligible only with hindsight.
 Therefore, we will only sketch the final result,  i.e. the complex $(\boldsymbol{\Var}^{*}_{r}, E_{i})$, as far as needed and present it in terms of the projections 
$
\bOme^k (J^{r-1} \bY) \mapsto \boldsymbol{\Var}^{*}_{r}
$; the interested reader will find the details of the construction in \cite{Kru15}.
For $0 \leq k \leq n$ we have
$$
\boldsymbol{\Var}^{k}_{r} := h (\boldsymbol{\Omega}^k (J^{r-1} \bY)) 
$$
where $h$ is the horizontalization. The differentials for $1 \leq k \leq n-1$ are then defined by
$$ 
E_{k}: \boldsymbol{\Var}^{k}_{r} \rightarrow \boldsymbol{\Var}^{k+1}_{r}  \qquad  E_{k} \left(h(\gamma)\right) = h(d \gamma)
$$
In addition we set
\beq
\boldsymbol{\Var}^{0}_{r} := C^{\infty} (J^{r-1} \bY)
\eeq
and
$$ 
E_{0}: C^{\infty} (J^{r-1} \bY) \rightarrow \boldsymbol{\Var}^{1}_{r}  \qquad  E_{0} (f) = h(df)
$$
$E_k$, $0 \leq k \leq n$, is sometimes also denoted by $d_h$ to underscore that it is just the horizontalization of the exterior derivative. The $\lambda \in \boldsymbol{\Var}^{n}_{r}$ are the {\bf Lagrangians} in the variational sequence.

For $k \geq n+1$ all $k$-forms are contact. To construct $\boldsymbol{\Var}^{k}_{r}$ for $k \geq n+1$ we need, thus, to distinguish contact forms which are ``essential'' for the calculus of variations from those which are ``inessential''. 

This is done by means of a family of operators $I_k$, $k \geq 1$, called the {\bf internal Euler operators}. $I_1 (\gamma)$ is characterized by the fact that the terms of its local expressions contain only the contact factors $\theta^i$ of first jet order and if we have locally,  for a $n+1$-form $\gamma$, the coordinate expression 
\beq
\gamma = \sum\limits_{i = 1}^{m} \left( \,A^i \cdot \theta^i + \, \sum\limits_{l = 1}^{r} \, \sum\limits_{\alpha_1 \dots \alpha_l} \, A^{i}_{\alpha_1 \dots \alpha_l} \cdot \theta^i_{\alpha_1 \dots \alpha_l} \wedge dx^1 \wedge \dots \wedge dx^n \right)
\eeq
then $I_1(\gamma)$ is defined by
\bEq \label{interEuler}
I_1 (\gamma)  = 
\sum\limits_{i = 1}^{m} \left( A^i + \sum\limits_{l = 1}^{r} \, \sum\limits_{\alpha_1 \dots \alpha_l} \, (-1)^l  d_{\alpha_1} \dots d_{\alpha_l} \, A^{i}_{\alpha_1 \dots \alpha_l} \right) \cdot \theta^i \wedge \omega^0 
\eEq
Here we have set $\omega^0 := dx^1 \wedge \dots \wedge dx^n$ and $d_{\alpha_l}$ is the total derivative with respect to the coordinate $x^{\alpha_l}$, see equation (\ref{totder}) for its local expression.

$I_k(\gamma)$ is a $k$-contact $n+k$-form for which each term of any of its local expressions contains at least one contact factor $\theta^i$ of first jet order.
All further information can be found in \cite{Kru15}. For our purposes only $I_1$ will be relevant. We set then for $k\geq1$
$$
\Var^{n+k}_{r} := I_k \circ p_k \left( \bOme^k (J^{r-1} \bY) \right) \qquad \Var^{n+1}_{r} \sub \bOme^{n+1} (J^{2r-1} \bY) 
$$
Note that there is an integer  $M:= m \left(\stackrel{n+r-1}{n}\right) + 2n - 1$ such that for $n+k > M$ we have $\Var^{M+l}_{r} = \bOme^{M+l} (J^{r-1} \bY)$ for $l \geq 1$. 
For $\lambda \in \Var^{n}_{r}$ and $\rho \in \bOme^n (J^{r-1} \bY)$ with $\lambda = h(\rho)$ (note that no matter if we start with $\lambda$ or $\rho$ there is always a pedant to complete the pair), we set then
\bEq \label{EuLa}
E_n (\lambda):= I_1 (d\lambda) = I_1 (d h(\rho)) =  I_1 \circ p_1 (d\rho) \qquad E_n :\Var^{n}_{r} \rightarrow \Var^{n+1}_{r}
\eEq
$E_n$ is called the {\bf Euler--Lagrange operator}.
Analogously, for $\gamma \in \bOme^{n+k} (J^{r-1} \bY)$ with $1\leq k \leq M-n-1$, we set
$$
E_{n+k}(I_k \circ p_k (\gamma)):= I_{n+k+1}(dp_{k}(\gamma)) \qquad E_n :\Var^{n+k}_{r} \rightarrow \Var^{n+k+1}_{r}
$$
For $\gamma \in \bOme^{M} (J^{2r-1} \bY)$, we have $E_{M} (I_M \circ p_M (\gamma)) = d\gamma$ and, obviously, for $i > M$ we have simply $E_i = d$. 
In the variational sequence formulation of the calculus of variations $E_{n} (\lambda)$ is the {\bf Euler--Lagrange form} and $E_{n} (\lambda) \circ j^{2r-1} \sigma = 0$ are the {\bf Euler--Lagrange equations}.

To see that this construction actually provides a cohomological formulation of the calculus of variations, one simply compares the relevant forms and expressions  in $(\boldsymbol{\Var}^{*}_{r}, E_{i})$ with the corresponding forms and expressions arising from variational problems on fibre bundles; in particular, one checks that $E_{n} (\lambda) = E_{\lambda}$.

Starting from $(\bOme^* (J^{r-1} \bY), d)$ the $r$th-order Lagrangians $\lambda = h(\gamma) \in \Var^{n}_{r}$ can be at most polynomial in the $y^i_{\alpha_1 \dots \alpha_r}$. To include all $r$th-order Lagrangians one would have to start the construction from $(\bOme^* (J^{r} \bY), d)$.
But we are exclusively interested in the Lagrangians of Chern--Simons and Yang--Mills theories which are polynomial in the $y^i_{\alpha_1 \dots \alpha_r}$. So we will not go into this.

The variational sequence  clarifies also the concept of {\bf local variational problem} (see the end of section \ref{VarBun} above): For any $\eta \in \Var^{n+1}_{r}$ with $E_{n+1}(\eta) = 0$ and any open covering  $(U_i)_{i \in I}$, there is a collection of local lagrangians $\lambda_i \in \Var^{n}_{r}|_{U_i}$ such that $\eta|_{J^{2r-1} U_i} = E_{n}(\lambda_i)$. Every local variational problem arises in this way. Of course, by setting $\lambda_i := \lambda|_{U_i}$ again for some open cover $(U_i)_{i \in I}$ any variational problem is also canonically a local variational problem.

Local variational problems are, of course, very interesting in their own right whenever the cohomology class $[\eta]$ is non zero. But much of their importance arises in situations when a source form in principal admits a global Lagrangian: global Lagrangians may not be easy to find or have undesirable properties, systems of local Lagrangians may be anyhow better to work with or a specific system may be the standard choice for historical reason. Chern--Simons gauge theories are prime examples for all these cases. 
\subsection{The obstruction} \label{Obs}
One of the interesting features of variational cohomology is that it makes the importance of the real cohomology of $\bX$ and $\bY$ in the calculus of variations explicit. Somewhat surprisingly, this aspect seems not to have been explored very much. In part this may depend on the fact that it is quite hard to explicitly describe variational cohomology classes in terms of the de Rham cohomology of $\bX$ and $\bY$. Also, the folk theorem that for variational problems relevant in physics all the cohomology classes vanish may have had its impact. Anyhow, the following theorem apparently has been overlooked until recently (\cite{PaWi16}, see also \cite{PaRoWiMu16}); we give here a (slightly different) proof to illustrate the workings of the variational sequence.
\bTh \label{obstruction}
Let $\eta_\lambda$  be the dynamical form of a local variational problem on a fibre bundle $\, \bY \mapsto \bX$ with $dim \bY > dim \bX=n$ and let $\pi^{*}: H^{n}_{dR}(\bX) $ $\mapsto$ $ H^{n}_{dR}(\bY)$ be an isomorphism. Let $\Xi$ be a vertical vector field such that $E_n (\Xi \rfloor\eta_\lambda)$ $ = $ $0$. Then both $[\Xi \rfloor\eta_\lambda] \in H^{n}_{dR}(J^{2r-1}\bY)$ and $[(j^{2r-1}\sigma)^* (\Xi \rfloor\eta_\lambda)] \in H^{n}_{dR}(\bX)$ for an arbitrary section $\sigma: \bX \mapsto \bY$ are obstructions to the existence of (global) solutions; $(j^{2r-1}\sigma)^* (\Xi \rfloor\eta_\lambda) \in H^{n}_{dR}(\bX)$ is independent of the section.
\eTh 
\bPf
Since $\eta$ is a $1$-contact form in $\bOme^{n+1}_{J^{2r-1} Y}(Y)$, $\Xi \rfloor \eta$ is horizontal in $\bOme^{n}_{J^{2r-1} Y}(Y)$. Thus, we find a 
$$
\beta \in h^{-1}(\Xi \rfloor \eta_{\lambda}) \subset \bOme^{n}_{J^{2r-2} Y}(Y)
$$
 with $d\beta = 0$ and which represents the same cohomology class as  $\Xi \rfloor \eta_{\lambda}$.
Since $(\pi^{2r-1}_{2r-2})^*\beta = \Xi \rfloor \eta_{\lambda} + \theta$ ($\theta$ is a contact $n$-form), we have 
$$
j^{2r-2}\sigma^*\beta = j^{2r-1}\sigma^*(\Xi \rfloor \eta_{\lambda})
$$
If we denote by $[\gamma]$ the cohomology class corresponding to the closed differerntial form $\gamma$ we have  
$$
j^{2r-1}\sigma^* [\Xi\rfloor \eta_{\lambda}] = [j^{2r-1}\sigma^*(\Xi \rfloor \eta_{\lambda})] = [j^{2r-2}\sigma^*\beta] = j^{2r-2}\sigma^*[\beta]
$$
Since $H^{n}_{dR}(\bY) \sim  H^{n}_{dR}(J^{2r-1}\bY)$ via the jet bundle projection, any jet prolongation of a section induces an inverse isomorphism to $(\pi^{k})^*$:
$$
(j\sigma^{k})^* \circ (\pi^{k})^* = 1_{H^{n}_{dR}(\bX)}
$$ 
and 
$$
(\pi^{k})^* \circ (j\sigma^{k})^* =1_{H^{n}_{dR}(J^* \bY)}
$$
Thus, $(j^{2r-1}\sigma)^* (\Xi \rfloor\eta_\lambda) \in H^{n}_{dR}(\bX)$ does not dependent on the section. Therefore, if $[\Xi \rfloor \eta_{\lambda}] \neq 0$ then also $[j^{2r-1}\sigma^*(\Xi \rfloor \eta_{\lambda})] \neq 0$ for all sections. Hence, $j^{2r-1}\sigma^*(\Xi \rfloor \eta_{\lambda})$ does not vanish along any section.

From the Euler--Lagrange equations $\eta_{\lambda} \circ j^{2r-1} \sigma = 0$ we see, however, that if $\sigma$ is a solution also $j^{2r-1}\sigma^*(\Xi \rfloor \eta_{\lambda}) = 0$. Therefore, if $[\Xi \rfloor \eta_{\lambda}] \neq 0$ there can be no solutions.
\ePf

There is a more general version of this theorem, see (\cite{PaWi16}). The cohmology class $[\Xi\rfloor \eta_{\lambda}]$ arises naturally in the context of the {\bf Noether-Bessel-Hagen theorem}, a kind of combination of the two Noether theorems (\cite{Noe18}) in variational cohomology. It is an obstruction to existence of global conserved quantities (see e.g. \cite{FePaWi10}, \cite{PaWi16}, also \cite{PaRoWiMu16}). 

General results regarding these obstructions seem virtually impossible to come by: they are defined via contractions and contractions behave badly with cohomology. Note that in the case of $ 0 = [\eta_{\lambda}] \in H_{dR}^{n+1}( \bY)$ we may still have $0 \neq [\Xi\rfloor \eta_{\lambda}] \in H^{n}_{dR}(\bX)$; in particular, this is true for Chern--Simons theories.

One word of warning: one cannot work exclusively with $(j^{2r-1}\sigma)^* (\Xi \rfloor\eta_\lambda)$. One needs $\Xi\rfloor \eta_{\lambda}$ to be closed in the variational sequence.

\section{The bundle of connections} \label{ConBun}
Yang--Mills and Chern--Simons gauge theories are a classical field theories for principal connections on a principal bundle $\bP \mapsto \bX$ over some $n$--dimensional smooth manifold $\bX$. To work with the variational sequence, it is therefore necessary to describe principal connections as sections of a bundle. So we need to introduce the bundle of connections (see also \cite{CaMun01} and the references therein).

A principal connection can be defined as decomposition of $T\bP$, the tangent bundle of $\bP$, into a direct sum of the vertical bundle $V\bP$ and a horizontal complement $\mathcal{H}$ which is invariant under the right action of $\bG$ on $\bP$, i.e.
$$
T\bP = \mathcal{H} \oplus V\bP  \qquad  \mathcal{H}_{yg} = dR_g (\mathcal{H}_{y})
$$
with $y \in \bP$, $g \in \bG$ and $dR_g$  the differential of the right action $R_g$. A second possibility is to define a principal connection as a right invariant Lie algebra valued one form $\omega$ on $\bP$ which maps the fundamental vector fields on their corresponding Lie algebra elements, i.e.
$$
\omega: T\bP \rightarrow \mathfrak{g} \qquad \omega(\widetilde{A}) = A \qquad \omega (dR_g (X)) = ad(g^{-1}) \omega (X)
$$
with $\mathfrak{g}$ the Lie algebra of $\bG$, $\widetilde{A} \in V\bP$, the fundamental vector field corresponding to $A \in \mathfrak{g}$, and $X \in T\bP$.
Or one can define a principal connection as a collection of $\mathfrak{g}$-valued one forms $\omega_U$ relative to a open cover $\mathcal{U}$ of $\bX$ over which $\bP$ can be trivialized, i.e.
$$
\omega_U : TU \mapsto \mathfrak{g} \qquad U \in \mathcal{U} \qquad \bP|_U \sim U \times \bG
$$
such that 
\bEq \label{trans}
\left(\omega_{U}(x) \right)|_{ U \cap V } \, (X) = \left(ad(h_{UV}(x)^{-1})(\omega_{V}(x)) + h_{UV}^{-1}(x)dh_{UV}(x) \right)|_{ U \cap V } \, (X)
\eEq
with $U, V \in \mathcal{U}$, $x \in U \cap V$ and $h_{UV}: U \cap V \mapsto \bG$ the transition function of the change of trivialisation from $\bP|_V$ to $\bP|_U$.

These different definitions are related by $\mathcal{H} = ker \, \omega$ and $\omega_{U} = (\gamma|_U)^* \omega$ with $\gamma|_U$ a section of $\bP|_U$ over $U \in \mathcal{U}$. Regarding the local coordinate expressions below, note that each $\gamma|_U$ defines a local trivialization $U \times \bG \sim \bP|_U$ by $ (x, g) \mapsto \gamma|_U(x) \cdot g$ and with respect to this trivialization we have $\gamma|_U(x) = (x, 1_{\bG})$. 
The bundle of connections of $\bP$  may now be defined by
\beq
\mathcal{C}_{\bP} := J^{1}\bP/\bG \mapsto \bP/\bG \sim \bX
\eeq
The sections $\sigma$ of this bundle are in one to one correspondence with the right invariant decompositions of $T\bP$ above by virtue of the canonical  decomposition of $(\pi^{1})^* T\bP$, see (\ref{candec1}) and the beginning of section \ref{candec}.
 $\pi_{\mathcal{C}_{\bP}}: \mathcal{C}_{\bP} \mapsto \bX$, is an {\bf affine bundle} modelled on the vector bundle 
\beq
T^* \bX \ten  V\bP/\bG \mapsto \bX,
\eeq
i.e. the bundle of $V\bP/\bG$ valued $1$-forms on $\bX$.

$\mathcal{C}_{\bP}$ inherits its affine structure from $J^{1}\bP$. It reflects the fact that the difference of two connections is a $V\bP/\bG$ valued $1$-form on $\bX$, but the connections themselves are not.
Note also that, since $\mathcal{C}_{\bP}$ is a (strong) deformation retract of $\bX$, we have $H^{*}_{dR}(\mathcal{C}_{\bP}) \sim  (H^{*}_{dR}(\bX))$ via $\pi^{*}$ and $\sigma^*$ for any section $\sigma$ of $\mathcal{C}_{\bP}$.
Locally $\mathcal{C}_{\bP}$ can be trivialized as $\mathcal{C}_{\bP}|_{U} \sim U \times \bR^{n} \ten \mathfrak{g}$ with $\mathfrak{g}$ the Lie algebra of $\bG$. If $(x^{\mu})_{1 \leq i \leq n}$ are coordinates on $\bX$ and $\mathfrak{e}_{i}$ is a base of  $\mathfrak{g}$, we have coordinates $(x^{\mu}, A^{\stackrel{i}{\mu}})$ ($A^{\stackrel{i}{\mu}}$ is the coefficient of the component $dx^{\mu} \ten \mathfrak{e}_{i}$) on $\mathcal{C}_{\bP}|_{U}$. The transition functions of $\mathcal{C}_{\bP}$ take the form
$$
k_{VU} (x) = ad(h_{UV}(x)^{-1}) \otimes J_{UV}^* + h_{UV}^{-1}(x)dh_{UV}(x) 
$$
i.e. the transition functions derive from the change of trivialization formula of the local connection forms \eqref{trans} and they are affine in view of the additive ``displacement term'' $ h_{UV}^{-1}(x)dh_{UV}(x)$; here $J_{UV}^*: T^*V|_{U\cap V} \mapsto T^*U|_{U\cap V}$ indicates the pullback with the change of coordinate Jacobian $J_{UV}: TU|_{U\cap V} \mapsto TV|_{U\cap V}$.

The contact structure of $J^{1}\bP$ defines the {\bf canonical connection} on the principal bundle  $J^{1}\bP \mapsto \mathcal{C}_{\bP}$ (see \cite{CaMun01}, note, however, that we use a different sign convention); at the point $q \in J^{1}\bP$ with coordinates $(x^{\mu}, A^{\stackrel{i}{\mu}}, g)$  with respect to a local trivialization $\mathcal{C}_{\bP}|_{U} \times \bG \sim U \times \bR^{n}\ten \mathfrak{g} \times \bG$ the connection form $\phi$ of this connection can be written as
\beq 
\phi_{(x^{\mu}, A^{\stackrel{i}{\mu}}, g)}  = ad(g^{-1})\left(\sum\limits_{i} \, \mathfrak{e}_{i} \ten (dg_{i} - \sum\limits_{\mu} \, A^{\stackrel{i}{\mu}} dx^{\mu}) \right) 
\eeq
$dg_{i}$ is defined at the point $q$ by $ \left(\mathfrak{e}_{i} \ten dg_{i} \right)_{q} \left( \, \widetilde{\mathfrak{e}}_{i} (q) \, \right) = \mathfrak{e}_{i}$, here 
$\widetilde{\mathfrak{e}}_{i}$ is the fundamental vector field corresponding to $\mathfrak{e}_{i}$; restricted to the fibre $\bG$ of $J^{1}\bP \mapsto \mathcal{C}_{\bP}$, $\sum\limits_i \mathfrak{e}_{i} \ten dg_{i}$ is, of course, the Maurer--Cartan form of $\bG$. 

Let $\gamma|_U: \mathcal{C}_{\bP}|_{U} \mapsto J^{1}\bP|_{U}$ be the section defined by $\gamma|_U(x, A^{\stackrel{i}{\mu}}) = (x, A^{\stackrel{i}{\mu}}, 1_{\bG})$.
For the local connection form $\phi_U$ we have then 
\beq 
\left(\phi_U\right)_{(x^{\mu}, A^{\stackrel{i}{\mu}})} = \left(\phi \circ \gamma|_U \right)_{(x^{\mu}, A^{\stackrel{i}{\mu}})} = - \sum\limits_{i} \, \, \mathfrak{e}_{i} \, \ten  \, \sum\limits_{\mu} \, A^{\stackrel{i}{\mu}} dx^{\mu} 
\eeq
$\omega_{\sigma}:= \phi \circ \sigma_{\bP}$ is the connection one form corresponding to the principal connection defined by the section $\sigma: \bX \mapsto \mathcal{C}_{\bP}$, where $\sigma_{\bP}$ is the lift of $\sigma$ to $\bP$ defined by the following  commutative diagram 
$$
\begin{CD}  
\\
\bP  @> \sigma_{\bP} >> J^1 \bP
\\
@V VV @V VV  
\\
\bX @> \sigma >> \mathcal{C}_{\bP}
\end{CD}
$$
The {\bf curvature} of a connection $\omega$ is 
$
\Omega = d\omega + \frac{1}{2}[\omega, \omega]
$.
Because of the right invariance of $\omega$ its curvature can be interpreted as either a $\mathfrak{g}$ valued two form on $\bP$ or a $V\bP / \bG$ valued two form on $\bX$. 
For the curvature $\mathcal{F}$ of $\phi$ we have the following local expression
\beq
\mathcal{F} = \sum\limits_{k} \, \mathfrak{e}_{k} \ten \, \left( \, \sum\limits_{\mu, \nu, \kappa} \, \left( dx^{\mu} \wedge dA^{\stackrel{k}{\mu}} + \sum\limits_{i, j} \, \frac{1}{2}c^{k}_{ij} \, A^{\stackrel{i}{\nu}}A^{\stackrel{j}{\kappa}} \, dx^{\nu} \wedge dx^{\kappa} \right) \right)
\eeq
($c^{k}_{ij}$ are the structural constants of $\mathfrak{g}$). 

For the horizontalization of $\mathcal{F}$ on $J^{1}\mathcal{C}_{\bP}$ we have then
\bEq \label{horF}
h(\mathcal{F}) = \sum\limits_{k} \, \mathfrak{e}_{k} \ten \, \left( \, \sum\limits_{\mu, \nu} \, \left( (A^{\stackrel{k}{\mu}}_{\,\, \nu} - A^{\stackrel{k}{\nu}}_{ \,\,\, \mu}) \, dx^{\mu} \wedge dx^{\nu}  + \sum\limits_{i, j} \, \frac{1}{2}c^{k}_{ij} \, A^{\stackrel{i}{\mu}}A^{\stackrel{j}{\nu}} \, dx^{\mu} \wedge dx^{\nu} \right) \right)
\eEq
 and we have the decomposition
$$
(\pi_0^1)^*\mathcal{F} = \Theta_{\phi} + h(\mathcal{F})
$$
where $\Theta_{\phi}$ is the contact component of $(\pi_0^1)^*\mathcal{F}$; locally 
\bEq \label{theta}
\Theta_{\phi} =  \sum\limits_{k}\,\, \mathfrak{e}_{k} \ten \, \sum\limits_{\mu} \,  dx^{\mu} \wedge \theta^{\stackrel{k}{\mu}} 
\eEq
($\theta^{\stackrel{k}{\mu}} = dA^{\stackrel{k}{\mu}} - \sum\limits_{\nu} \, A^{\stackrel{k}{\mu}}_{\,\, \nu} \, dx^{\nu}$ are the local contact one forms on $J^{1}\mathcal{C}_{\bP}$).

Depending on the interpretation of $\mathcal{F}$  we have either 
$$
\Theta_{\phi}, h(\mathcal{F}) \in \Lambda^2 (J^{1}\mathcal{C}_{\bP}) \ten V\bP / \bG 
$$
 or 
$$
\, \Theta_{\phi}, h(\mathcal{F}) \in \Lambda^2 ( (\pi_0^1)^* J^{1}\bP) \ten \mathfrak{g}
$$
$(\pi_0^1)^* J^{1}\bP$ is the total space of the pullback of the principal bundle $J^{1}\bP \mapsto \mathcal{C}_{\bP}$ to $J^{1}\mathcal{C}_{\bP}$ via the jet bundle projection $\pi_0^1: J^{1}\mathcal{C}_{\bP} \mapsto \mathcal{C}_{\bP}$

Finally, for the curvature $\Omega_{\sigma}$ of the connection $\omega_{\sigma}$ with $\sigma(x) = \left(x, A^{\stackrel{k}{\mu}}(x) \right)$ locally, we have 
\bEq \label{Omega}
\Omega_{\sigma} = (j^{1}\sigma)^{*}h(\mathcal{F})  = \sigma^{*} \mathcal{F} 
\eEq 
and the coordinate expression
\beq
\Omega_{\sigma} = 
\eeq 
{\footnotesize \beq  
\sum\limits_{k} \, \mathfrak{e}_{k} \ten \, \left( \, \sum\limits_{\mu, \nu} \, \left( \left(\frac{\der}{\der x_\nu}A^{\stackrel{k}{\mu}} (x) - \frac{\der}{\der x_\mu}A^{\stackrel{k}{\nu}}(x)\right) \, dx^{\mu} \wedge dx^{\nu}  + \sum\limits_{i, j} \, \frac{1}{2}c^{k}_{ij} \, A^{\stackrel{i}{\mu}}(x)A^{\stackrel{j}{\nu}}(x) \, dx^{\mu} \wedge dx^{\nu} \right) \right) 
\eeq}

\section{Chern--Simons theories on the bundle of connections} \label{CSBun}

Chern--Simons theories are gauge theories for connections. Starting point is the work by Chern and Simons on secondary characteristic classes, \cite{ChSi71} and \cite{ChSi74}.  

Let $\cP_{k}$ be an ad$G$-invariant polynomial of degree $k$ on $\mathfrak{g}$ the Lie algebra of $\bG$, let $\omega$ be a connection on the principal bundle $\bP$ over $\bX$ and $\Omega$ its curvature form
$\mathcal{P}_{k}(\Omega)$ is a closed form of degree $2k$ on $\bP$. Horizontal and invariant under the right action of $G$, it can also be considered a closed form on $\bX$. The cohomology classes thus defined are independent of the connection $\omega$ (this is the starting point of the {\bf Chern--Weil theory} of characteristic classes). 

Before continuing, we will introduce some operations on Lie algebra-- or matrix--valued forms and the conventions with which we use them. Let $\mathfrak{g}$ be a Lie algebra over $\bR$ or $\bC$, $(\mathfrak{e}_i)_{i \in I}$ a basis of it and $\alpha$ and $\beta$ $\bR$-- or $\bC$--valued differential forms. We set now
\beq
[\mathfrak{e}_i \ten \alpha, \mathfrak{e}_j \ten \beta] := [\mathfrak{e}_i, \mathfrak{e}_j] \ten \alpha \wedge \beta
\eeq
and
\beq
\mathfrak{e}_i \ten \alpha \, \wedge \, \mathfrak{e}_j \ten \beta := \left(\mathfrak{e}_i \ten \mathfrak{e}_j\right) \, \ten \, \alpha \wedge \beta
\eeq
and extend the {\bf Lie bracket} $[,]$ and the {\bf exterior product} $\wedge$ by linearity to arbitrary $\mathfrak{g}$--valued differential forms. Note that the Lie bracket of $\mathfrak{g}$--valued forms is $\mathfrak{g}$--valued, while their exterior product takes values in $\mathfrak{g} \ten \mathfrak{g}$. 

The product of matrices $A = (\alpha_{ij})_{1 \leq i,j \leq n}$ and $B = (\beta_{ij})_{1 \leq i,j \leq n}$ of differential forms is the matrix of differential forms defined by
\beq
AB \, = \, C \qquad C  = (\gamma_{ij})_{1 \leq i,j \leq n}
\eeq 
with
\beq
\gamma_{ij} \,\,  = \,\, \sum\limits_{k = 1}^n \, \alpha_{ik} \wedge \beta_{kj}
\eeq
To not overload the notation even further we will allow the following ambiguity for the expression $\Upsilon^i$ for a $\mathfrak{g}$--valued form $\Upsilon$
\begin{itemize}
\item wherever $\mathfrak{g}$ appears as an abstract Lie algebra, e.g. when dealing with invariant polynomials on $\mathfrak{g}$ in general terms, we will set $\Upsilon^i = \Upsilon \wedge ... \wedge \Upsilon$ ($i$-fold wedge product)
\item if instead $\Upsilon$ is given  with respect to a specific representation of $\mathfrak{g}$, $\Upsilon^i$ will be the $i$-fold matrix product of $\Upsilon$ with itself
\end{itemize}
Chern and Simons found in \cite{ChSi74} a $2k-1$-form  $T\mathcal{P}_{k}(\omega)$ on $\bP$ such that 
\beq 
\mathcal{P}_{k}(\Omega) = dT\mathcal{P}_{k}(\omega)
\eeq
on $\bP$. One can write explicitly (\cite{ChSi74})
\beq 
T\mathcal{P}_{k}(\omega) = \sum\limits_{i=0}^{k-1} \,  \frac{(-1)^i \cdot k! \cdot (k-1)!}{2^i \cdot (k+i)! \cdot (k-1-i)!} \cdot \mathcal{P}_{k}(\omega \wedge [\omega, \omega]^{i} \wedge \Omega^{k-i-1})
\eeq
If $\mathcal{P}_{k}(\Omega)=0$ we have $0= dT\mathcal{P}_{k}(\omega)$ and, thus, $T\mathcal{P}_{k}(\omega)$ defines a cohomology class on $P$ which may depend on the connection $\omega$. These cohomology classes are the {\bf Chern--Simons secondary characteristic classes}. Expository material about Chern--Weil theory, Chern--Simons secondary characteristic classes and their applications is easily accessible, so only a few classics: \cite{Ch95}, \cite{ChSi71}, \cite{ChSi74}, chapter XII and appendix $20$ of Volume II of \cite{KobNo}.

If dim $\bX = 2p+1$, $T\mathcal{P}_{2p+2}(\omega)$ depends on the connection. The central idea behind Chern--Simons gauge theories is now to use this dependence  to construct a local variational principle for connections. From now on $\bP \mapsto \bX$ will be a $\bG$-principal bundle over the $(2p+1)$-dimensional manifold $\bX$.
On $J^{1}\bP$ we have  
$$
dT\mathcal{P}_{p+1}(\phi) = \mathcal{P}_{p+1}(\mathcal{F})
$$ 
for the canonical connection $\phi$ on the principal bundle  $J^{1}\bP \mapsto \mathcal{C}_{\bP}$. 

Let $\mathcal{U }$ be now an open cover of $\mathcal{C}_{\bP}$. Using a family of local section $\alpha_U: U \mapsto J^{1}\bP|_U \,\, U \in \mathcal{U}$ , we get a system of local connection forms $\phi_U := \phi \circ \alpha_U$. Thus, on $\mathcal{C}_{\bP}$ we have locally
$$
dT\mathcal{P}_{p+1}(\phi_{U})  = \mathcal{P}_{p+1}(\mathcal{F})|_U \,.
$$
The horizontalization of these local potentials of $\mathcal{P}_{p+1}(\mathcal{F})$ is now the system of local Lagrangians  on $J^{1}\mathcal{C}_{\bP}$ of the Chern--Simons theory 
\beq
\lambda^{CS}_{U} := h \left(T\mathcal{P}_{p+1}(\phi_{U}) \right) = \sum\limits_{i=0}^{p} \, \kappa_{i} \cdot \mathcal{P}_{p+1}\left(\phi_{U} \wedge [\phi_{U}, \phi_{U}]^{i} \wedge \left( h (\mathcal{F})|_{U}\right)^{p-i}\right)
\eeq
The Euler-Lagrange form of this local variational problem is (see (\ref{EuLa}) above)
\beq
\eta_{CS} = I_1 \circ p_1 \left(\mathcal{P}_{p+1}(\mathcal{F})\right ) = I_1 \left(\mathcal{P}_{p+1}(\Theta_{\phi} \wedge h(\mathcal{F})^p)\right) = \mathcal{P}_{p+1}(\Theta_{\phi} \wedge h(\mathcal{F})^p)
\eeq
Note that the last equality holds because $\mathcal{P}_{p+1}(\Theta_{\phi} \wedge h(\mathcal{F})^p)$ contains only first order contact terms. We summarize this:
\bPr \label{CStheory}
Let $\bP \mapsto \bX$ be a $\bG$ principal bundle over the $(2p+1)$-dimensional manifold $\bX$; let $\mathcal{P}_{p+1}$ be an $ad(\bG)$-invariant polynomial of degree
 $p+1$ on $\mathfrak{g}$, the Lie algebra of $\bG$. On an open cover $\mathcal{U}$ of the respective bundle of connections $\mathcal{C}_{\bP} \mapsto \bX$ the system of local Lagrangians 
\beq
\lambda^{CS}_{U} :=  \sum\limits_{i=0}^{p} \kappa_{i}\mathcal{P}_{p+1}\left(\phi_{U} \wedge [\phi_{U}, \phi_{U}]^{i} \wedge \left( h (\mathcal{F})|_{U}\right)^{p-i}\right)
\eeq
defines a first order local variational problem, the Chern--Simons theory of $(\bP, \mathcal{P}_{p+1})$. Its Euler-Lagrange form is 
\beq
\eta_{CS} = \mathcal{P}_{p+1}(\Theta_{\phi} \wedge h(\mathcal{F})^p) 
\eeq
\ePr

Note that in this case not only the Lagrangian, but also the Euler-Lagrange form is of first order, i.e. defined on $J^1 \mathcal{C}_{\bP}$.

Since $H^{n+1} (J^{r}\mathcal{C}_{\bP}) \sim H^{n+1} (\mathcal{C}_{\bP}) \sim H^{n+1} (X) = 0$, Chern--Simons  theories are in principle globally variational. But a global Lagrangian seems to depend always on fixing a physical quantity, a background connection, {\em a priori} (see \cite{BFF03} for the $3$-dimensional case and \cite{GiMaSa03} for higher dimensions). Anyhow, the questions of what is the correct or best Chern--Simons Lagrangian is completely irrelevant to us, since we will occupy ourselves exclusively with cohomological invariants extracted from the Euler--Lagrange form. 

\section{Unitary Chern--Simons theories and Yang--Mills--Chern--Simons theories} \label{Char}
We will now turn to unitary Chern--Simons theories derived from the polynomial $ch_{p+1}$; the polynomial $ch_{k}$ corresponds via Chern--Weil theory to the $k$th component of the Chern character. 

We will introduce and analyze a specific obstruction of type $\Xi \rfloor \eta_{CS}$ and apply it in the context of the theory of solitons (instantons) in Yang--Mills--Chern--Simons theories, i.e. Lagrangian gauge field theories obtained adding Yang--Mills and Chern--Simons Lagrangians.

At present, Yang--Mill--Chern--Simons theories seem to be under consideration as physical theories exclusively in five dimensions, in holographic QCD. However, in view of the long standing and ongoing interest in higher dimensional gauge theories in theoretical physics and mathematics (see e.g. \cite{DS11}, \cite{DT99}, \cite{GKS84}, \cite{NST16}, \cite{Naka17}, \cite{Take17}, \cite{Tian00}), we will work in arbitrary dimensions until we deal explicitly with holographic QCD.
\subsection{Chern--Simons theories for $U(n)$ or $SU(n+1)$, \\ $(n \geq p)$, and the polynomial $ch_{p+1}$} \label{Char2}
We proceed now with the analysis in the case of the $k$th component of the Chern character in a similar way as with the theories derived from the $(p+1)$th Chern class in \cite{PaWi22}. The the components of the Chern character and the Chern classes are derived from two different sets of invariant polynomials which are related by the Newton identities (see e.g. \cite{ChW76}).

 $\bP \mapsto \bX$ will now  be a  $U(n)$ or $SU(n+1)$ principal bundle over the $(2p+1)$-dimensional manifold $\bX$ with $(n \geq p)$. As before, all matrix expressions are relative to the standard representations of $U(n)$ and $SU(n+1)$ and their Lie algebras  $\mathfrak{u}_{n}$ or $\mathfrak{su}_{n+1}$. 
For an arbitrary matrix $A$ denote by $tr A$ its trace; we define then an invariant polynomial $ch_{k}$ on $(\mathfrak{s})\mathfrak{u}_{n(+1)}$
\bEq \label{charpol}
ch_{k} (X) = \left(\frac{i}{2\pi}\right)^{k} tr X^{k} \qquad X \in (\mathfrak{s})\mathfrak{u}_{p(+1)}
\eEq
If $\Omega$ is the curvature $2$-form of a connection $\omega$ on $\bP$, we have by Chern--Weil theory
\bEq \label{chernchar}
[ch_{k}(\Omega)] = ch_{k}(\bP) \in H^{2k} (\bX, \R)
\eEq
where $ch_{k}(\bP)$ is the $k$th component of the Chern character of $\bP$. In algebraic topology, the Chern character is a formal power series in the Chern classes; $\left(\frac{i}{2\pi}\right)^{k}$ is a normalization factor such that $[ch_{k}(\Omega)]$ lies in the image of $H^{2k} (\bX, \Z)$ in $H^{2k} (\bX, \R)$ induced by the inclusion $\Z \subset \R$.
By proposition \ref{CStheory}, we have then
\beq
ch_{p+1}(\Theta_{\phi} + h(\mathcal{F})) =  \left(\frac{i}{2\pi}\right)^{p+1} tr (\Theta_{\phi} + h(\mathcal{F}))^{p+1} 
\eeq
and 
\beq 
\eta_{CS} = ch_{p+1}(\Theta_{\phi} \wedge h(\mathcal{F})^p) =  \left(\frac{i}{2\pi}\right)^{p+1} tr \left(\Theta_{\phi} \, h(\mathcal{F})^p \right) = 
\eeq
\bEq \label{chareq}
\left(\frac{i}{2\pi}\right)^{p+1} \cdot \sum\limits_{k}\sum\limits_{j} \left(\Theta_{\phi}\right)_{kj} \cdot \left(h(\mathcal{F})^p\right)_{jk}
\eEq
for the Chern--Simons gauge field theory derived from $ch_{p+1}$; note that the latter two expressions are again globally well defined on $(\pi_0^1)^* J^{1}\bP$.
\bPr \label{eqchar}
Let $\bP \mapsto \bX$ be a  $U(n)$ or $SU(n+1)$ principal bundle over the $(2p+1)$-dimensional manifold $\bX$, $(n \geq p)$; let $\omega_{\sigma}$ be the principal connection on $\bP$ corresponding to the section $\sigma$ of $\pi_{\mathcal{C}_{\bP}}: \mathcal{C}_{\bP} \mapsto \bX$ and let $\Omega_{\sigma}$ be its curvature. $\sigma$ is then a solution of the Chern--Simons gauge field theory derived from $ch_{p+1}$ if and only if
\beq
(\Omega_{\sigma})^p = 0
\eeq
($p$-fold matrix product of $\Omega_{\sigma}$ with itself taken with respect to the standard representation of $(\mathfrak{s})\mathfrak{u}_{n(+1)}$).
\ePr
\bPf
We have to find the conditions under which $ \eta_{CS} \, \circ \, j^1 \sigma = 0$. The key observation is that $\Theta_{\phi}$ does not vanish along any section (see equation (\ref{theta})) In particular, none of the $\left(\Theta_{\phi}\right)_{kj}$ vanish along sections. Thus, equation (\ref{chareq}) implies that all the $\left(h(\mathcal{F})^p\right)_{jk}|_{\sigma} = 0$. Since $\Omega_{\sigma} = (j^{1}\sigma)^{*}h(\mathcal{F})$, see equation (\ref{Omega}), this means that all the $(\Omega_{\sigma})^p_{jk}$ vanish (everywhere locally), i.e. $(\Omega_{\sigma})^p = 0$.
\ePf

For a solution $\omega_{\sigma}$ this implies, of course, $ch_p(\Omega_{\sigma}) = tr(\Omega_{\sigma})^p = 0$ and, thus, $0 = [ch_{p}(\Omega_{\sigma})] = ch_{p}(\bP) \in H^{2k} (\bX, \R)$. Hence, $ch_{p}(\bP)$ is an obstruction to the existence of solutions of the Euler--Lagrange equations.

For the applications to Yang--Mills--Chern--Simons theories in the next section,  we will now construct an obstruction of type $\Xi \rfloor \eta_{CS}$ that vanishes if and only if $ch_{p}(\bP)$ vanishes.

For the rest of this section and for the non existence results of the next (see theorem \ref{noso}) will restrict ourselves to closed manifolds of dimension $2p+1$. This compactness hypothesis serves to streamline the exposition somewhat and clarify the result. 
In section \ref{QCD} we will discuss its adaption to the non compact case. 

We note that on a closed ($2p+1$--dimensional) manifold $\bX$ there exists, by Poincar\`e duality, a closed one form $\beta$
such that $0 \neq [\beta \wedge ch_{p}(\Omega_{\sigma})] \in H^{2p+1}(\bX, \R)$ whenever $0 \neq [ch_{p}(\Omega_{\sigma})] \in H^{2p}(\bX, \R)$. This means, we need to find a vertical vector field $\Xi$ on $\mathcal{C}_{\bP}$ such that
\beq
j^1 \sigma^* \left(\Xi \rfloor \eta_{CS}\right) = \beta \wedge ch_{p}(\Omega_{\sigma})
\eeq
First we will deal with the case of an $U(n)$--principal bundle $\bP \mapsto \bX$, $n \geq p$. 
Since $\pi_{\mathcal{C}_{\bP}}: \mathcal{C}_{\bP} \mapsto \bX$, is an affine bundle modeled on the vector bundle 
\beq
T^* \bX \ten  V\bP/U(n) \mapsto \bX
\eeq
we have
\beq
V\mathcal{C}_{\bP} \, \sim \, \mathcal{C}_{\bP} \, \times_{\bX} \, T^* \bX \ten V\bP /U(n)
\eeq
and, in particular, we can identify
\beq
\frac{\der}{\der A^{\stackrel{k}{\mu}}} \, = \, \mathfrak{e}_{k} \ten dx^{\mu}
\eeq
Since $i \cdot  \bf{1_{n}} \in \mathfrak{u}(n)$ is $ \, ad \,U(n)$--invariant, $- i \cdot  \bf{1_{n}} \ten \alpha$
is then identified with a global vertical vector field for any differential one form $\alpha$ on $\bX$ and we have by equation (\ref{chareq})
\beq
j^1 \sigma^* \left(( - i \cdot  \bf{1_{n}} \ten \alpha) \rfloor \eta_{CS}\right) \, = \, \frac{-1}{2\pi} \cdot \left(\frac{i}{2\pi}\right)^{p} \cdot \alpha \wedge tr(\Omega_{\sigma}^p)  \, = \, \frac{-1}{2\pi} \cdot \alpha \wedge ch_{p}(\Omega_{\sigma})
\eeq
In particular, $j^1 \sigma^* \left(( - i \cdot  \bf{1}_{p} \ten \alpha) \rfloor \eta_{CS}\right)$ is closed if and only if $\alpha$ is and  we can choose $\alpha$ to be the Poincar\`e dual $\beta$ above if $0 \neq [ch_{p}(\Omega_{\sigma})]$. 
It remains to be shown that $( - i \cdot  \bf{1_{n}} \ten \alpha) \rfloor \eta_{CS}$ is closed whenever $j^1 \sigma^* \left(( - i \cdot  \bf{1_{n}} \ten \alpha) \rfloor \eta_{CS}\right)$ is. This follow immediately from the fact that
\beq
\frac{-1}{2\pi} \cdot \alpha \wedge ch_{p}(\Omega_{\sigma}) = \sigma^* \left( \frac{-1}{2\pi} \cdot \pi^* (\alpha) \wedge ch_{p}(\mathcal{F}) \right)
\eeq
and 
\beq
( - i \cdot  \bf{1_{n}} \ten \alpha) \rfloor \eta_{CS} = h\left( \frac{-1}{2\pi} \cdot \pi^* (\alpha) \wedge ch_{p}(\mathcal{F})\right)
\eeq
i.e. if $j^1 \sigma^* \left(( - i \cdot  \bf{1_{n}} \ten \alpha) \rfloor \eta_{CS}\right)$ is closed, then $( - i \cdot  \bf{1_{n}} \ten \alpha) \rfloor \eta_{CS}$ is the horizontalization of a closed differential form  on $\mathcal{C}_{\bP}$ and, therefore, closed  by construction of the variational sequence, see section \ref{VarCo}.

For a $SU(n+1)$--principal bundle $\bP \mapsto \bX$ ($n \geq p$), we note that the inclusion $SU(n+1) \subset U(n+1)$ induces the inclusion $\bP\subset \bP_{U(n+1)}$, with $\bP_{U(n+1)} \sim \bP  \times_{SU(n+1)} U(n+1)$. $\bP_{U(n+1)}$ is then a $U(n+1)$--principal bundle. For the respective bundles of connections this means:
\beq
\mathcal{C}_{\bP_{U(n+1)}} \, \sim \, \mathcal{C}_{\bP} \, \times_{\bX} \, \mathcal{C}_{\bU(1)}  \, \sim \, \mathcal{C}_{\bP} \, \times_{\bX} \, \R
\eeq
Hence,
\begin{itemize}
\item every $SU(n+1)$--principal connection on $\bP$ extents canonically to a $U(n+1)$--principal connection on $\bP_{U(n+1)}$ by adding the canonical flat connection on $\bU(1)$ 
\item the Chern--Simons gauge field theory derived from $ch_{p+1}$ on $\mathcal{C}_{\bP}$ extents obviously to $\mathcal{C}_{\bP_{U(n+1)}}$; however, only the $U(n+1)$--principal connections on $\bP_{U(n+1)}$ with $\mathfrak{u}(1)$-component the canonical flat connections are the ``fields'' of the $SU(n+1)$-Chern--Simons theory
\item a solution of the Euler--Lagrange equations on $\mathcal{C}_{\bP}$ is also a solution of the Euler--Lagrange equations on $\mathcal{C}_{\bP_{U(n+1)}}$
\end{itemize}
Thus, $( - i \cdot  \bf{1_{n}} \ten \alpha) \rfloor \eta_{CS}$, respectively $j^1 \sigma^* \left(( - i \cdot  \bf{1}_{p} \ten \alpha) \rfloor \eta_{CS}\right)$, is also a obstruction in the $SU(n+1)$ case. We summarize this:

\bPr \label{obstruction4}
Let $\bP \mapsto \bX$ be a  $U(n)$ or $SU(n+1)$ principal bundle over the closed $(2p+1)$-dimensional manifold $\bX$, $(n \geq p)$ and let $\eta_{CS}$ be the Euler-Lagrange form of the Chern--Simons gauge field theory derived from $ch_{p+1}$. Let $\alpha$ be a closed one form on $\bX$ and let furthermore $\Xi_{\alpha}$ be the vertical vector field on $\mathcal{C}_{\bP}$ corresponding to $- i \cdot  \bf{1_{n}} \ten \alpha$. 

The cohomology classes $[\Xi_{\alpha} \rfloor \eta_{CS}] \in H^{2p+1}(J^1\mathcal{C}_{\bP}, \R)$ and $[j^1 \sigma^* \left(\Xi_{\alpha} \rfloor \eta_{CS}\right)] \in H^{2p+1}(\bX, \R)$ are then obstructions to the existence of solutions of the Euler--Lagrange equations.
These obstructions vanish for all $\Xi_{\alpha}$ if and only if $0 = ch_{p}(\bP) \in H^{2p+1}(\bX, \R)$.
\ePr
Regarding the $SU(n+1)$ case: that $ch_{p}(\bP)$ is an obstruction, is already implicit in the Euler--Lagrange form, see proposition \ref{eqchar}. The point of the above extension to the $U(n+1)$-case is that it expresses this fact in terms of the obstructions $[\Xi_{\alpha} \rfloor \eta_{CS}]$. 
For ''coupled'' Lagrangians this allows to identify the precise counterterms which would need to annihilate the obstructions, as we will see in the next section for Yang--Mills--Chern--Simons theories. For $SU(n+1)$ theories one will have to keep track of effect of the vanishing of the $\mathfrak{u}(1)$-component of the curvature on these counterterms.

\subsection{On the non existence of solitons in  Yang--Mills--Chern--Simons theories} \label{nosoliton}
We will now explicate the how the obstructions $\Xi_{\alpha} \rfloor \eta_{CS}$ of proposition \ref{obstruction4} essentially impede the existence of solitons in Yang--Mills--Chern--Simons theories. Adopting fairly common terminology (see e.g. \cite{ManSut04}), by ``topological soliton'' or simply ``soliton'' we will refer to any solution of the Euler--Lagrange equations of a (locally) Lagrangian quantum field theory. The solitons of Yang--Mills theory in four dimensions are, of course, usually called ``instantons''. 
The literature about Yang--Mills theory and instantons is extraordinary vast, so we will give only some pointers.
As general introduction to  Yang--Mills theories may serve \cite{MaMa92}. As introduction to instantons we refer to \cite{Atiyah79}, \cite{AHS78} and \cite{Cole79}. A good survey of instantons in theoretical physics is \cite{ABC99}. And, finally, for the r\^ole of instantons in differential geometry, see \cite{DK97} in four dimensions and \cite{DS11}, \cite{DT99} and \cite{Tian00} for the possibility of extending these ideas to higher dimensions.

Yang--Mills--Chern--Simons theories are (locally) Lagrangian field theories on odd dimensional manifolds, the Lagrangian of which is a sum  of the Yang--Mills and Chern--Simons Lagrangians:
\beq
\lambda_{YMCS} = \lambda_{YM} + \kappa \cdot \lambda_{CS} 
\eeq
($\kappa$ is a constant factor, the so called coupling constant). This construction seems to motivated by the r\^ole of the Chern--Simons form in anomaly physics (see e.g. \cite{AGG85}, \cite{BaZu84} and \cite{Zum84}); for holographic QCD this is made explicit e.g. in \cite{Hill06a} and \cite{Hill06b}. There the idea is roughly as follows. The Yang--Mills--Chern--Simons theory is defined on $\R^5$ (with Lorentzian metric). This $\R^5$ is the interior of a manifold with boundary $\bM$ with the boundary $\der\bM$ considered to be at ``infinity''. The point of Yang--Mills--Chern--Simons theories is then that the Chern--Simons term causes anomaly cancellation in the Yang--Mills theory on the boundary; an anomaly is a conserved quantity corresponding to a symmetry of the classical theory which is not a symmetry of the quantum theory. 
Such relations between Field theories on the interior (physicists seem to call this the ``bulk'') and their restrictions to the boundary of a manifold is what physicists call ``holography''; the boundary is also often referred to  as ``holographic''. Apparently section $2$ of \cite{Wit98} was the first explicit description of such an interplay between field theories on the bulk and on the boundary. 

The Chern--Simons Lagrangians are, by construction, only locally defined and, thus, is $\lambda_{YMCS}$. But since Chern--Simons theories are local variational problems (see section \ref{CSBun}), also Yang--Mills--Chern--Simons theories are. 
The corresponding Euler--Lagrange form (on $J^2 \mathcal{C}_{\bP}$) is
\beq
\eta_{YMCS} = E_n (\lambda_{YM}) + \kappa \cdot (\pi^2_1)^* \eta_{CS} 
\eeq
Let $\bX$ be an oriented Riemannian manifold or an oriented Pseudo-Riemannian manifold of Lorentzian signature and let $\bP$ be a $U(n)$--principal bundle over $\bX$. 
The unitary Yang--Mills theory on $J^1\mathcal{C}_{\bP}$ is  the Lagrangian field theory  with Lagrangian
\beq
\lambda_{YM} = <h(\mathcal{F}), h(\mathcal{F})> d_{m}\bX
\eeq
where $d_{m}\bX$ is the metric volume form on $\bX$ and $<,>$ is induced in the usual way by the (Pseudo-)Riemannian metric on $\bX$ and the $ad U(n)$-invariant metric on $\mathfrak{u}_n$ defined by $<A, B> = - tr(AB)$, with $A, B \in \mathfrak{u}_n$ in the standard representation. 
Of course, this is completely equivalent to the usual formulation of unitary Yang--Mills theories on the (affine) space of connections $\mathcal{A}$. 
$\mathfrak{u}_n$ being canonically isomorphic to $\mathfrak{u}_1 \oplus \mathfrak{su}_n$, every principal connection $\omega$ on $\bP$ and its curvature $\Omega$, as well as the canonical connection $\phi$ on $J^1 \bP \mapsto \mathcal{C}_{\bP}$ and its curvature $\mathcal{F}$ decompose into a sum of an $\mathfrak{u}_1$ and a $\mathfrak{su}_n$ component. For the Yang--Mills Lagrangian we have then
\beq
\lambda_{YM} = <h(\mathcal{F}_{\mathfrak{u}_1}), h(\mathcal{F}_{\mathfrak{u}_1})> d_{m}\bX + <h(\mathcal{F}_{\mathfrak{su}_n}), h(\mathcal{F}_{\mathfrak{su}_n})> d_{m}\bX
\eeq
Hence, by linearity,  also its Euler--Lagrange form decomposes into an $\mathfrak{u}_1$ and a $\mathfrak{su}_n$ component
\beq
E_n (\lambda_{YM}) = E_n (\lambda_{YM})_{\mathfrak{u}_1} + E_n (\lambda_{YM})_{\mathfrak{su}_n}
\eeq
Note that the existence of a Pseudo-Riemannian metric of Lorentzian signature poses no restriction an orientable $2p+1$-dimensional manifold $\bX$, since such a manifold has always an everywhere non zero vector field. This is because the only obstruction to the existence of such a vector field is the Euler class of $\bX$, i.e. of its tangent bundle, and the Euler class of an odd dimensional vector bundle is of order $2$, while $H^{2p+1} (\bX, \Z) \sim 0$ if $\bX$ is non compact and $H^{2p+1} (\bX, \Z) \sim \Z$ if $\bX$ is closed; thus, the Euler class of $\bX$ vanishes always.
Let now $\bX$ be again of dimension $2p+1$ and closed. 

$\Xi_{\alpha}$ is the vertical vector field on $\mathcal{C}_{\bP}$ corresponding to $- i \cdot  \bf{1_{n}} \ten \alpha$ as in proposition \ref{obstruction4}. Since in the standard representation $\mathfrak{u}_1 = - i \, r \cdot  \bf{1_{n}}$ (with $ r \in \R$), we have 
\beq
j^2 \sigma^* \left(\Xi_{\alpha} \rfloor E_n (\lambda_{YM})\right) = j^2 \sigma^* \left(\Xi_{\alpha} \rfloor E_n (\lambda_{YM})_{\mathfrak{u}_1}\right)  = 
j^2 \sigma^* \left(\Xi_{\alpha} \rfloor I_1 \left( d(\lambda_{YM})_{\mathfrak{u}_1}\right) \right) = 
\eeq
\beq
j^2 \sigma^* \left(\Xi_{\alpha} \rfloor I_1 \left(\,\, d<h(\mathcal{F}_{\mathfrak{u}_1}), h(\mathcal{F}_{\mathfrak{u}_1})> \wedge \,\, d_{m}\bX\right) \right) = 
\alpha \wedge d*(\Omega_{\sigma})_{\mathfrak{u}_1}
\eeq
where $*$ is the Hodge star operator with respect to the the Pseudo-Riemannian or Riemannian metric on $\bX$. The last equality can be derived fairly easily from the local expressions for the internal Euler operator, see equation (\ref{interEuler}), $h(\mathcal{F}_{\mathfrak{u}_1})$, cf. equation (\ref{horF}), and $(\Omega_{\sigma})_{\mathfrak{u}_1}$, cf. equation (\ref{Omega}). Together with the general expression for a metric on $\bX$, the resulting formulae become, however, quite formidable to write down. We leave working out the details to the interested reader. 
This equality must hold also on general grounds, since $E_n (\lambda_{YM})_{\mathfrak{u}_1}$ is the Euler--Lagrange form of electromagnetism and $d*(\Omega_{\sigma})_{\mathfrak{u}_1} = 0$ are the source-free Maxwell equations in $2p+1$ dimensions. 
For the Euler--Lagrange form of the Yang--Mills--Chern--Simons theory we have then
\beq
j^2 \sigma^* \left(\Xi_{\alpha} \rfloor \eta_{YMCS} \right) = j^2 \sigma^* \left(\Xi_{\alpha} \rfloor E_n (\lambda_{YM}) \right) 
+ \kappa \cdot j^2 \sigma^* \left(\Xi_{\alpha} \rfloor (\pi^2_1)^* \eta_{CS} \right) =
\eeq
\beq
 j^2 \sigma^* \left(\Xi_{\alpha} \rfloor E_n (\lambda_{YM})_{\mathfrak{u}_1} \right) 
+ \kappa \cdot j^2 \sigma^* \left(\Xi_{\alpha} \rfloor (\pi^2_1)^* \eta_{CS} \right) =
\eeq
\beq
\alpha \wedge d*(\Omega_{\sigma})_{\mathfrak{u}_1} + \frac{-\kappa}{2\pi} \cdot \left(\frac{i}{2\pi}\right)^{p} \cdot \alpha \wedge ch_{p}(\Omega_{\sigma})
\eeq
For a hypothetical solution $\sigma$ of the Euler--Lagrange equations we would have $j^2 \sigma^* \left(\Xi_{\alpha} \rfloor \eta_{YMCS} \right) = 0$ and, thus,
\bEq \label{YMCS}
\alpha \wedge d*(\Omega_{\sigma})_{\mathfrak{u}_1} = \frac{\kappa}{2\pi} \cdot \left(\frac{i}{2\pi}\right)^{p} \cdot \alpha \wedge ch_{p}(\Omega_{\sigma})
\eEq
For $SU(n+1)$-theories (a $SU(n+1)$-theory can always be viewed as an $U(n+1)$-theory, see section \ref{Char2} above) this is equivalent to 
\beq
\frac{\kappa}{2\pi} \cdot \left(\frac{i}{2\pi}\right)^{p} \cdot \alpha \wedge ch_{p}(\Omega_{\sigma}) = 0
\eeq
and this is , of course, possible if and only if $0 = ch_{p}(\bP) \in H^{2p}(\bX, \R)$.  
In the case of $U(n)$-theories, since $\alpha$ is closed, we have 
\beq
\alpha \wedge d*(\Omega_{\sigma})_{\mathfrak{u}_1} =  - d\left(\alpha \wedge *(\Omega_{\sigma})_{\mathfrak{u}_1}\right)
\eeq
Thus, the Yang--Mills component of $j^2 \sigma^* \left(\Xi_{\alpha} \rfloor \eta_{YMCS} \right)$ is always a coboundary and cohomologically trivial. For the Chern--Simons component, instead, exists always a vector field $\Xi_{\alpha}$ such that it is cohomologically non trivial as long as $0 \neq ch_{p}(\bP) \in H^{2p}(\bX, \R)$ (see section \ref{Char2} above). 

This means that equation \ref{YMCS} cannot have solutions if  $0 \neq ch_{p}(\bP) \in H^{2p}(\bX, \R)$. As a consequence, the vanishing of $ch_{p}(\bP) \in H^{2p}(\bX, \R)$ is a necessary condition for the Euler--Lagrange equations $j^2 \sigma \circ \eta_{YMCS} = 0$ to have solutions. Hence, we have the following non existence theorem.
\bTh \label{noso}
Let $\bP$ be a $U(n)$-- or $SU(n+1)$--principal bundle over the closed, $2p+1$-dimensional oriented Riemannian or oriented Pseudo-Riemannian manifold $\bX$ of Lorentzian signature, $n \geq p$. Let also $0 \neq ch_{p}(\bP) \in H^{2p}(\bX, \R)$. 

The Euler--Lagrange equations $\eta_{YMCS} \circ j^2 \sigma = 0$ of the corresponding unitary Yang--Mills--Chern--Simons theory on $J^1\mathcal{C}_{\bP}$ do not admit solutions.
\eTh

For Yang--Mills theories on $4$-dimensional Riemannian manifolds, $\int ch_{2}(\Omega)$ is up to a numerical factor the instanton number. Since this integral vanishes if and only if $ch_{2}(\bP)$ is cohomologically trivial, this indicates, by analogy, how restrictive this non existence theorem is. However, we will not define a ``soliton number'' for Yang--Mills--Chern--Simons theories at the present level of generality. But we will address the issue in the next section for the Yang--Mills--Chern--Simons theories of $5$-dimensional QCD
\subsubsection{On the existence of solitons in the Yang--Mills--Chern--Simons theories of $5$-dimensional (holographic) QCD} \label{QCD}
We will now apply the ideas of section \ref{nosoliton} above to the Yang--Mills--Chern--Simons theories of $5$-dimensional (holographic) QCD. Solitons are a central part in many of these theories. They represent a particular class of particles called baryons on which relies the treatment of nuclear matter, see e.g. \cite{KMS12,KMS15,KoS20,RSZ10}.

Probably the most important among the YMCS theories of holographic QCD is the Sakai--Sugimoto model, see \cite{SaSu1,SaSu2}. Its solitons have been extensively studied, see e.g. \cite{BaBo17,BoSu14,HSS07,HSSY07,HRY07}. 

Constructions of such Yang--Mills--Chern--Simons theories start usually with an $\R^5$ with coordinates $x_0, x_1, x_2, x_3, z$, where the $x_i$ are the coordinates of Minkowski spacetime, $x_0$ being time, and $z$ is the so called holographic coordinate, and equipped with a not necessarily flat Pseudo-Riemannian metric $\mu$ of signature $(-, +, +, +, +)$ (or $(+, -, -, -, -)$), see e.g. \cite{BaBo17,BoSu14}.

This $\R^5$ (or some subset of it, defined by so called infrared and ultraviolet cut offs somewhere on the holographic axis) is then the Pseudo-Riemannian base manifold $\bX$ of a principal bundle $\bP$ on which the Yang--Mills--Chern--Simons theories will be set up. One considers mainly theories with group $U(n)$, but also $SU(n+1)$ (see e.g. \cite{Hill06a}), $n \geq 2$. 

In analogy with the instantons of $4$--dimensional Yang--Mills theory, the solitons of holographic QCD are required to satisfy the finiteness condition that the integral
\bEq \label{BN}
\tau \cdot \int_{x_0 = const.} ch_{2}(\Omega)
\eEq
converges (see e.g. \cite{BoSu14}); here, $\Omega$ is the curvature of an principal connection $\omega$ on the unitary principal bundle $\bP \mapsto \R^5$ and $\tau$ is a numerical factor. 
For this $\Omega$ needs, of course, to go sufficiently fast to $0$ for $r \mapsto \infty$ with $r = \sqrt{x_1^2 + x_2^2 + x_3^2 + z^2}$ . 

The most natural and mathematical consistent way (as well as apparently the only one ever used systematically) to guarantee that the integral (\ref{BN}) is well defined, is to require that $\bP \mapsto \R^5$ is the restriction of a principal bundle $\widehat{\bP} \mapsto \R \times S^4$ and the connections $\omega$ are the restrictions of the principal connections $\widehat{\omega}$ on $\widehat{\bP}$. The restriction is with respect to the embedding $s_+: \R^5 \mapsto \R \times S^4$ which is defined by being the identity on the coordinate $x_0$ and, say, the stereographic projection $\R^4 \mapsto S^4$ (with respect to $(0, 0, 0, 0, 1) \in \R^5$) on $(x_1, x_2, x_3, z)$. 

$\R \times S^4$ can now be covered by $s_+$ and $s_-$ which is defined analogously to $s_+$ but using the opposite stereographic projection, i.e. the one with respect to $(0, 0, 0, 0, -1) \in \R^5$, instead. Every smooth $U(n)$-- principal bundle, $n \geq 2$, $\widehat{\bP} \mapsto \R \times S^4$ is then isomorphic to 
\bEq \label{bunstr}
\left(\R^5_+ \times U(n) \cup \R^5_- \times U(n)\right) /_{\sim} \, \mapsto \, \left(\R^5_+ \cup \R^5_- \right) /_{\sim} 
\eEq
for $x \in \R^5_+$ and $y \in \R^5_-$ the latter equivalence relation is simply that $x \sim y$ if and only if $ x = s_+^{-1}( s_-(y))$. The former instead is that $(x, g) \sim (y, h)$ if and only if $x \sim y$ and $g = \nu_{+-}(x)h$ with a smooth map $\nu_{+-}: \R \times (\R^4 - 0) \mapsto U(n)$. 

The isomorphism classes are then in one to one correspondence with the homotopy classes of the maps $\nu_{+-}$. Since $\R \times (\R^4 - 0)$ can be contracted onto $S^3$, the set of isomorphism classes can be identified with $\pi_3 \left( U(n) \right) \sim \Z$. In particular, we can choose $\nu_{+-}$ to be constant along the $x_0$-direction and along the radial direction in $\R^4 - 0$, i.e. 
\bEq \label{constnu}
\nu_{+-} = \tilde{\nu}_{+-} \circ \pi^{\R \times (\R^4 - 0)}_{S^3}
\eEq
with $\tilde{\nu}_{+-}: S^3 \mapsto S^3$ and $\pi^{\R \times (\R^4 - 0)}_{S^3}: \R \times (\R^4 - 0) \mapsto S^3$ is the projection along the radial and $x_0$-direction. From now on $\nu_{+-}$ will be of this type.

Pulling back the (positive) generator of $H^3 \left( U(n), \Z \right)$ with $\nu_{+-}$ one can identify the set of isomorphism classes also with $H^3 \left( \R \times (\R^4 - 0), \Z \right) \sim H^3 \left( S^3, \Z \right) \sim \Z$. 
Note that the inclusion $\Z \subset \R$ induces an inclusion $H^3 \left( \R \times (\R^4 - 0), \Z \right) \subset H^3 \left( \R \times (\R^4 - 0), \R \right)$, since $H^3 \left( \R \times (\R^4 - 0), \Z \right)$ is torsion free. 
The isomorphism classes can then be represented by the forms 
\bEq \label{3form}
\kappa \cdot tr\left(\phi [\phi, \phi]\right)
\eEq
here, $\phi = \nu_{+-}^{-1}d\nu_{+-}$ is the pullback of the Maurer--Cartan form of $U(n)$ by $\nu_{+-}$ and $\kappa$ is the usual normalization factor such that the cohomology class is in $H^3 \left( \R \times (\R^4 - 0), \Z \right) \subset H^3 \left( \R \times (\R^4 - 0), \R \right)$.

Via the connecting homomorphism in the Mayer-Vietoris sequence for $\R^5_+$, $\R^5_-$ and $\R \times S^4 \sim \left(\R^5_+ \cup \R^5_- \right) /_{\sim}$ the set of isomorphism classes can also be identified with $H^4 \left( \R \times S^4, \Z \right) \sim H^4 \left( S^4, \Z \right) \sim \Z$; $ch_{2}(\widehat{\bP})$ can then be interpreted as the cohomology class which represents the isomorphism class of $\widehat{\bP}$. 

Again, the inclusion $\Z \subset \R$ induces an inclusion $H^4 \left( \R \times S^4, \Z \right) \subset H^4 \left( \R \times S^4, \R \right)$, since $H^4 \left( \R \times S^4, \Z \right)$ is torsion free. 

The numerical factor $\tau$ can be chosen such that the integral (\ref{BN}) takes integer values and since $[ch_{2}(\Omega)] = ch_{2}(\widehat{\bP}) \in H^{4}(\R \times S^4, \R)$, the integral can then be viewed as the identification $H^4 \left( \R \times S^4, \Z \right) \sim \Z$. 
Thus, also the values of the integral (\ref{BN}) classify the $U(n)$-- principal bundles over $\R \times S^4$ up to isomorphism. This value is then identified as the {\bf baryon number} and is the analog of the instanton number of $4$--dimemsional Yang--Mills theory.

Regarding the physical meaning, for us the following fairly naive interpretation will suffice. Baryons are the particles of which matter is made. They are composed of three quarks ``[...] and a “sea” of quark-antiquark pairs'' (\cite{PaN7} p. 253). The respective antiparticles are antibaryons with three antiquarks instead of the three quarks. In physics, the baryon number is then defined as a third of the difference of the number of quarks and antiquarks.  We won't bother with specific particles and their names, but apparently it is a lot more complicated than just protons and neutrons (\cite{PaN7} chapter 16).

The solitons of Yang--Mills--Chern--Simons theories of $5$-dimensional QCD are interpreted as (multi-) baryons or antibaryons and the value of the integral (\ref{BN}) indicates the number of baryons present if it is positive and the number of anti baryons if it is negative. 

Technically, this is done by adapting the construction by Atiyah and Manton of deriving skyrmions in $3$ dimensions from instantons in $4$ dimensions by calculating the holonomy along lines parallel to the time axis. In this way the solitons in $5$ dimensions produce ``holographic'' skyrmions, i.e defined on the holographic boundary of the $5$-dimensional theory mentioned at the beginning of section \ref{nosoliton}, and these holographic skyrmions are then the model for baryons in holographic QCD (see \cite{Sut15} for an extensive review of  holographic skyrmions; see \cite{AtMan89,ManSut04} for the original construction by  Atiyah and Manton). 

Our observations, however, do not touch upon this, but are only concerned with the existence of solutions for the $5$-dimensional theory, i.e. we can and will continue to ignore the holographic boundary, albeit it may be the most important feature of holographic QCD.

That the baryon number be not only finite, but also conserved is a fundamental requisite of particle physics (\cite{PaN7} chapter 8.1). Here, this is guaranteed by the fact that different baryon numbers belong to topologically non isomorphic principal bundles and, therefore, there is no way to deform one into the other. A way to meet these requirements without the asymptotic behaviour at infinity imposing this topologically non trivial situation is yet to be found. 

Thus, mathematical consistency would require that only solitons on $\R^5$ wich extend to solitons on $\R \times S^4$ are admissible for ``producing'' holographic skyrmions. 

\bRm  \label{important}
In all this it is, of course, completley irrelevant that $s_+$ is the the stereographic projection.  
Any other $q_+: \R^5 \mapsto \R \times S^4$ which is the identity on the $x_0$--coordinate and embedds $\R^4$  (the coordinates $(x_1, x_2, x_3, z)$) diffeomorphically into $S^4$ will do. The crucial point is that the connection and curvature forms are pulled back from a principal bundle on $\R \times S^4$. This will guarantee then that their coefficients decrease fast enough for the integral \ref{BN} to converge and that the {\bf baryon number} will be conserved.
\eRm
For the study of baryonic matter, of course, only the cases with non zero baryon number are relevant. Thus, $0 \neq ch_{2}(\widehat{\bP}) \in H^{4}(\R \times S^4, \R)$ and $\widehat{\bP}$ is, therefore, non trivial.

As a special case of theorem \ref{noso} we know from equation \ref{YMCS} that solitons with non zero baryon number do not exist on a closed, oriented Pseudo-Riemannian $5$-manifold of Lorentzian signature. We will now discuss what changes in the present setting where our YMCS-theory is defined on the non compact manifold $\R^5$ as a restriction of a  YMCS-theory on $\R \times S^4$ via the embedding $s_+: \R^5 \mapsto \R \times S^4$, see above.

Let then $\widehat{\bP} $ be a smooth $U(n)$-- principal bundle, $n \geq 2$, over $\R \times S^4$ with $ch_{2}(\widehat{\bP})$ cohomologically non trivial and let $\Xi$ be the vertical vector field on $\mathcal{C}_{\bP}$ corresponding to $dx_0 \ten (- i \cdot \bf{1}_{n})$. Equation (\ref{YMCS}) becomes then on the $\R^5 \subset \R \times S^4$

\bEq \label{YMCS2}
dx_0  \wedge d*(\Omega_{\sigma})_{\mathfrak{u}_1} = \frac{\kappa}{2\pi} \cdot \left(\frac{i}{2\pi}\right)^{p} \cdot dx_0  \wedge ch_{2}(\Omega_{\sigma})
\eEq
For a $k$--form $\alpha$ on $\R^5$ we define 
$\alpha_{\bx}$ by 
$
\frac{\der}{\der_{x_0}} \, \rfloor \, \alpha_{\bx} = 0
$
and
$
dx_0 \wedge \alpha = dx_0 \wedge \alpha_{\bx}
$.
 We can then derive from equation (\ref{YMCS2})
\bEq \label{YMCS3}
\left(d*(\Omega_{\sigma}^0)_{\mathfrak{u}_1}\right)_{\bx} = \frac{\kappa}{2\pi} \cdot \left(\frac{i}{2\pi}\right)^{p} \cdot ch_{2}(\Omega_{\sigma})_{\bx}
\eEq
Let now $\varepsilon_t: \R^4 \mapsto \R^5$ be the  embedding which identifies $(x_1, x_2, x_3, z) \in \R^4$ with $(t, x_1, x_2, x_3, z) \in \R^5$. We have then
\beq 
\varepsilon_t^* \left(d*(\Omega_{\sigma}^0)_{\mathfrak{u}_1}\right)_{\bx} = \varepsilon_t^* d*(\Omega_{\sigma})_{\mathfrak{u}_1} = 
d\left( \varepsilon_t^* \left(*(\Omega_{\sigma})_{\mathfrak{u}_1}\right)\right) = d\left( \varepsilon_t^* \left(*(\Omega_{\sigma})_{\mathfrak{u}_1}\right)_{\bx}\right)
\eeq
and
\beq
\varepsilon_t^*ch_{2}(\Omega_{\sigma})_{\bx} = \varepsilon_t^*ch_{2}(\Omega_{\sigma})
\eeq
Thus on $\R^4$ equation (\ref{YMCS3}) becomes 
\bEq \label{YMCS4}
d\left( \varepsilon_t^* \left(*(\Omega_{\sigma})_{\mathfrak{u}_1}\right)_{\bx}\right) = \frac{\kappa}{2\pi} \cdot \left(\frac{i}{2\pi}\right)^{p} \cdot \varepsilon_t^*ch_{2}(\Omega_{\sigma})
\eEq
 
 We can draw now our first conclusions.
\bCr
Let $0 \neq ch_{2}(\widehat{\bP}) \in H^{4}(\R \times S^4, \R)$. In this case $SU(n)$-theories will not admit solitons.
For $U(n)$-theories instead, for the existence of solitons it would be necessary that not both the metric $\mu$ and $(\Omega_{\sigma})_{\mathfrak{u}_1}$ not extend to $\R \times S^4$.
\eCr 
\bPf
To have solutions of the Euler--Lagrange equations, it is necessary that equations (\ref{YMCS2}), (\ref{YMCS3}) and (\ref{YMCS4}) are satisfied. If $ch_{2}(\Omega_{\sigma})$ is cohomologically non trivial, so is $\varepsilon_t^*ch_{2}(\Omega_{\sigma})$. By Equation (\ref{YMCS4}) this means that $d\left( \varepsilon_t^* \left(*(\Omega_{\sigma})_{\mathfrak{u}_1}\right)_{\bx}\right)$ has to extend to $S^4$, but $*(\Omega_{\sigma})_{\mathfrak{u}_1}$ must not, else $d\left(\varepsilon_t^* \left(*(\Omega_{\sigma})_{\mathfrak{u}_1}\right)_{\bx}\right)$ would be cohomologically trivial on $S^4$. 
By the definition of $\varepsilon_t$ this means in turn that $\left(d*(\Omega_{\sigma}^0)_{\mathfrak{u}_1}\right)_{\bx}$ needs to extend to $\R \times S^4$, but $\left(*(\Omega_{\sigma}^0)_{\mathfrak{u}_1}\right)_{\bx}$ and, thus, $*(\Omega_{\sigma})_{\mathfrak{u}_1}$ must not.
But $*(\Omega_{\sigma})_{\mathfrak{u}_1}$ depends only on the metric $\mu$ and $(\Omega_{\sigma})_{\mathfrak{u}_1}$. Therefore, if both extend to $\R \times S^4$,  $*(\Omega_{\sigma})_{\mathfrak{u}_1}$ would  as well. 

For $SU(n)$-theories holds the same argument as in section \ref{Char2} above: without an  $\mathfrak{u}_1$-component of the curvature the left hand sides of equations (\ref{YMCS2}), (\ref{YMCS3}) and (\ref{YMCS4}) vanish and the right hand sides do not.
 \ePf

In particular, a non extending $(\Omega_{\sigma})_{\mathfrak{u}_1}$ would impose on the $\mathfrak{u}_1$-component of the curvature an ``asymptotic behaviour at infinity'' qualitatively different from the one of the $\mathfrak{su}_n$-component. Such a violation of the requirement that solitons extend to $\R \times S^4$ or any other inconsistenca imposed on the $\mathfrak{u}_1$-component can hardly be acceptable. 

The consequences of a non extending metric, on the other hand, are rather more subtle: it might cause the breakdown of the Yang--Mills part of the theory, because its Lagrangian and its Euler--Lagrange form might not extend as well. This is, however, by no means necessary: the metric of four dimensional Yang--Mills theory does not extend either, strictly speaking (it is, of course, conformally related to a metric which extends and the Yang--Mills Lagrangian is in four dimensions invariant under conformal changes of the metric).
Moreover, for holographic QCD the only relevant requirement is that the metric behaves well with the holographic boundary.

To gain a more precise understanding of the situation, we start by analyzing the the pullback to $\R^4$ restriction $ch_{2}(\Omega)$ to $\R^5$.
With our discussion of the classification of principal bundles over $\R \times S^4$ in mind we arrive then, e.g. from the construction of the connecting homomorphism in the Mayer--Vietoris sequence for de Rham cohomology, at the following relation between the forms (\ref{3form}) and the restriction $ch_{2}(\Omega)|_{\R^5}$
\bEq \label{charpot}
 \varepsilon_t^*ch_{2}(\Omega)|_{\R^5} =  d\left(\varepsilon_t^*\beta + \frac{r^2}{r^2 + 1}  \cdot \left(tr\left(\phi [\phi, \phi]\right) + d\alpha  \right) \right)
\eEq
where $r = \sqrt{x_1^2 + x_2^2 + x_3^2 + z^2}$; $\beta$ is a $3$-form on $S^4$; $\alpha$ is an arbitrary $2$-form on $\R^4 - \{ 0 \}$. 

By inserting equation (\ref{charpot}) into equation (\ref{YMCS4}) we can compare $\left(*(\Omega_{\sigma}^0)_{\mathfrak{u}_1}\right)_{\bx}$ and the local potential of $ch_{2}(\Omega)$:
\beq \label{charpoteq}
d\left(\varepsilon_t^* \left(*(\Omega_{\sigma})_{\mathfrak{u}_1}\right)_{\bx}\right) = \frac{\kappa}{2\pi} \cdot \left(\frac{i}{2\pi}\right)^{p}  \cdot  d\left(\varepsilon_t^*\beta + \frac{r^2}{r^2 + 1}  \cdot \left(tr\left(\phi [\phi, \phi]\right) + d\alpha  \right) \right) 
\eeq

Since $^{\lim}_{r \, \mapsto \infty} \, \varepsilon_t^*\beta = 0$, we have
\bEq \label{charpoteq2}
^{\lim}_{r \, \mapsto \infty} \, \left(*(\Omega_{\sigma})_{\mathfrak{u}_1}\right)_{\bx} \sim \delta \cdot tr\left(\tilde{\phi} [\tilde{\phi}, \tilde{\phi}]\right) + \,\,  ^{\lim}_{r \, \mapsto \infty} \, d\alpha
\eEq
here $\tilde{\phi} = \tilde{\nu}_{+-}^{-1}d\tilde{\nu}_{+-}$, see equation (\ref{constnu}) above, $\delta$ is a numerical normalization factor and $^{\lim}_{r \, \mapsto \infty} \, d\alpha$ is some exact $3$-form on the $3$--sphere ``at infinity``; i.e. $\delta \cdot tr\left(\tilde{\phi} [\tilde{\phi}, \tilde{\phi}]\right) + \,\,  ^{\lim}_{r \, \mapsto \infty} \, d\alpha$ is a differential form representing a cohomology class of the $3$--sphere ``at infinity`` to which $\left(*(\Omega_{\sigma})_{\mathfrak{u}_1}\right)_{\bx}$ has to converge for $r \, \mapsto \infty$.

In practice it is quite difficult to decide if \eqref{charpoteq2} is satisfied. To ilustrate this we will calculate $\left(*(\Omega_{\sigma})_{\mathfrak{u}_1}\right)_{\bx}$ in the case os the important Sakai--Sugimoto model.
The metric in the Sakai--Sugimoto model is 
\beq
\mu = \left(1 + \frac{z^2}{L^2}\right)^{\frac{2}{3}}\cdot \left( \, -(dx_0)^2 + \sum\limits_{i = 1}^3 (dx_i)^2 \, \right) \, + \,  \frac{1}{(1 + \frac{z^2}{L^2})^{\frac{2}{3}}} \cdot (dz)^2
\eeq
where $L$ is a positive constant which we normalize to $1$. If we define the coefficients of $(\Omega_{\sigma})_{\mathfrak{u}_1} - \left((\Omega_{\sigma})_{\mathfrak{u}_1}\right)_{\bx}$ by 
\beq
(\Omega_{\sigma})_{\mathfrak{u}_1} - \left((\Omega_{\sigma})_{\mathfrak{u}_1}\right)_{\bx} = 
\eeq
\beq
(\Omega_{\sigma})_{\mathfrak{u}_1}^{01} dx_0 \wedge dx_1 + (\Omega_{\sigma})_{\mathfrak{u}_1}^{02} dx_0 \wedge dx_2
+ (\Omega_{\sigma})_{\mathfrak{u}_1}^{03} dx_0 \wedge dx_3 + (\Omega_{\sigma})_{\mathfrak{u}_1}^{0z} dx_0 \wedge dz
\eeq
(note that in the Ansatz of \cite{BoSu14} we have $\left((\Omega_{\sigma})_{\mathfrak{u}_1}\right)_{\bx} = 0$), then we have 
\beq
\left(*(\Omega_{\sigma})_{\mathfrak{u}_1}\right)_{\bx} =
\eeq
\beq
\frac{-1}{(1+z^2)^\frac{1}{3}} \left((\Omega_{\sigma})_{\mathfrak{u}_1}^{01} dx_2 \wedge dx_3 \wedge dz 
- (\Omega_{\sigma})_{\mathfrak{u}_1}^{02} dx_1 \wedge dx_3 \wedge dz
+ (\Omega_{\sigma})_{\mathfrak{u}_1}^{03} dx_1 \wedge dx_2 \wedge dz \right)
\eeq
\bEq \label{SaSu}
+ \, (1+z^2)  (\Omega_{\sigma})_{\mathfrak{u}_1}^{0z} \, dx_1 \wedge dx_2 \wedge dx_3
\eEq
The ambiguties in $\delta \cdot tr\left(\tilde{\phi} [\tilde{\phi}, \tilde{\phi}]\right) + \,\,  ^{\lim}_{r \, \mapsto \infty} \, d\alpha$ ($\tilde{\phi}$ is defined only up to homotopy; the choice of the $2$-form $\alpha$ on $\R^4 - \{ 0 \}$ is completely arbitrary) render the comparison with the right hand side of \eqref{SaSu}, as required by \eqref{charpoteq2}, quite difficult. One may try and write the right hand side of \eqref{charpoteq2} differently, say using a multiple of the volume form of the $3$-sphere instead of  $\delta \cdot tr\left(\tilde{\phi} [\tilde{\phi}, \tilde{\phi}]\right)$, but the problems with verifying \eqref{charpoteq2} remain.

There are other models for Yang--Mills--Chern--Simons theories of $5$-dimensional QCD (see e.g. \cite{BRW10}, \cite{PoWu09}) which are quite similar. Unfortunately, the models of \cite{BRW10} and \cite{PoWu09} are constructed on subsets of $\R^5$ defined by the ``infrared'' and ``ultraviolet'' cutoffs and the exposition leaves it somewhat obscure how they are supposed to be embedded into $\R \times S^4$ or some $\R \times \bM$; cf. the discussion of equation (\ref{BN}) above. Therefore, checking \eqref{charpoteq2} becomes even more complicated.
Note, however, that the study of the solitons in all these theories is very much based on approximations or numerical methods.  Therefore, it seems indicated to show that the existencs of such solitons in a mathematically consistent way is guaranteed either via \eqref{charpoteq2} or by some other mathematically rigorous approach.

\section*{Acknowledgements}
Research partially supported by Department of Mathematics - University of Torino through the project PALM$\_$RILO$\_15\_ 01$ `Strutture geometriche e algebriche in fisica matematica e applicazioni'.
The Author wishes to thank M. Palese for useful discussions.



\begin{thebibliography}{99}

\bibitem{AGG85}  L.. Alvarez-Gaum\'e, P. Ginsparg: The Structure of Gauge and Gravitational Anomalies {\em Annals of Physics} {\bf 162} 
(1985) 423--490.

\bibitem{Atiyah79} M. F. Atiyah: Geometry of Yang--Mills fields, {\em Lezioni Fermiane} Accademia Nazionale dei Lincei - Scuola Normale Superiore,
 Pisa 1979 

\bibitem{AHS78} M. F. Atiyah, N. Hitchin, I. M. Singer: Self-duality in four-dimensional Riemannian geometry
{\em Proc. R. Soc. Lond. A.}
{\bf 362} (1978) 425--461

\bibitem{AtMan89}M. F. Atiyah, N. S. Manton: Skyrmions from instantons, {\em Phys. Lett.B} {\bf 222} (1989) 438--442

\bibitem{BaVeCa20} F. Bajardi, D.Vernieri and S. Capozziello: Exact Solutions in Higher-Dimensional Lovelock and $AdS_5$Chern–Simons Gravity, {\em JCAP} {\bf 11} 057 (2021) 
DOI 10.1088/1475-7516/2021/11/057.

\bibitem{BaBo17} S. Baldino, S. Bolognesi, S.B. Gudnason and D. Koksal,: Solitonic approach to holographic
nuclear physics {\em  Phys. Rev. D }{\bf 96} (2017) 034008

\bibitem{BaZu84} W. A. Bardeen and B. Zumino, Consistent and Covariant Anomalies in Gauge and
Gravitational Theories {\em Nucl. Phys. B} {\bf 244} (1984) 421--453

\bibitem{BRW10} D. Becciolini, M. Redi, A. Wulzer:  AdS/QCD:The Relevance of the Geometry {\em JHEP } {\bf 2010} (2010) 074

\bibitem{BoSu14} S. Bolognesi, P. Sutcliffe: The Sakai-Sugimoto soliton {\em JHEP }{\bf 2014} (2014) 078 

\bibitem{BFF03} A. Borowiec,  M. Ferraris,  M. Francaviglia: A covariant formalism for Chern--Simons gravity, {\em  J. Phys. }{\bf A 36} (10) (2003) 2589--2598. 

\bibitem{CaMun01} M.Castrillon Lopez, J. Munoz Masque: The geometry of the bundle of connections, {\em Math. Z.} {\bf 236} (2001) 797--811.

\bibitem{Ch95} S.S.Chern: Geometry of characteristic classes; in S. S. Chern: {\em Complex Manifolds without Potential Theory} Springer Verlag, Second Edition, NewYork 1995

\bibitem{ChSi71}  S.S. Chern, J. Simons: Some cohomology classes in principal fiber bundles and their application to Riemannian geometry, {\em Proc. Nat. Acad. Sci. U.S.A.} {\bf 68} (1971) 791--794. 

\bibitem{ChSi74} S.S. Chern, J. Simons: Characteristic forms and geometric invariants, {\em Ann. of Math.} {\bf (2) 99} (1974) 48--69. 

\bibitem{ChW76} S.S. Chern, J. White: Duality Properties of Characteristic Forms, {\em Inventiones math.} {\bf 35} (1976) 285--297. 

\bibitem{Cole79} S. Coleman: Uses of Instantons; in A. Zichichi Eds.: {\em The Whys of Subnuclear Physics}, Proceedings of the International School of Subnuclear Physics,
Ettore Majorana, Erice, 1977; Plenum, New York 1979

\bibitem{DK97} S. K. Donaldson, P. B. Kronheimer: The Geometry of Four-Manifolds; {\em Oxford Mathematical Monographs} Clarendon Press, Oxford 1997

\bibitem{DS11} S. K. Donaldson, E. Segal: Gauge theory in higher dimensions, II; in N. C. Leung, S. T. Yau Eds.: {\em Surveys in Differential Geometry XVI} International Press of Boston, Inc, Boston 2011

\bibitem{DT99} S. K. Donaldson, R. Thomas: Gauge theory in higher dimensions; in Huggett et.al. Eds.: {\em The
geometric universe: science, geometry, and the work of Roger Penrose} Oxford University Press, Oxford 1998

\bibitem{FePaWi10}  M. Ferraris, M. Palese, E. Winterroth: Local variational problems and conservation laws,  {\em Diff. Geom. Appl} {\bf 29} (2011) S80--S85. 

\bibitem{GiMaSa03} G. Giachetta, L. Mangiarotti, G. Sardanashvily: Noether conservation laws in higher-dimensional Chern--Simons theory, {\em Modern Phys. Lett. } {\bf A 18} (37) (2003) 2645--2651. 

\bibitem{GiMaSa00} G. Giachetta, L. Mangiarotti, G. Sardanashvily: Cohomology of the variational bicomplex on infinite order jet space,

\bibitem{GKS84} B. Grossman, T. W. Kephart, J. D. Stasheff: Solutions to Yang-Mills Field Equations in Eight Dimensions and the Last Hopf Map, {\em Commun. Math. Phys. } {\bf 96} (1984) 431--437.

\bibitem{HSS07} K. Hashimoto, T. Sakai and S. Sugimoto: Holographic baryons: static properties and form
factors from gauge/string duality {\em Prog. Theor. Phys.} {\bf 120} (2008) 1093

\bibitem{HSSY07} H. Hata, T. Sakai, S. Sugimoto, S. Yamato: Baryons from instantons in holographic
QCD, {\em Prog. Theor. Phys.} {\bf 117} (2007) 1157

\bibitem{Hill06a} C. Hill: Anomalies, Chern-Simons terms and chiral delocalization in extra dimensions, {\em Physical Review D } {\bf 73} (2006) 085001

\bibitem{Hill06b} C. Hill: Exact equivalence of the D = 4 gauged Wess-Zumino-Witten term and the D = 5 Yang-Mills
Chern-Simons term, {\em Physical Review D } {\bf 73} (2006) 126009

\bibitem{HRY07} D. K. Hong, M. Rho, H.-U. Yee, P. Yi: Chiral dynamics of baryons from string theory, {\em Phys. Rev. D} {\bf 76} (2007) 061901 

\bibitem{KMS12} V. Kaplunovsky, D. Melnikov, and J. Sonnenschein: Baryonic popcorn, {\em JHEP}
{\bf 1211} (2012) 047

\bibitem{KMS15} V. Kaplunovsky, D. Melnikov, and J. Sonnenschein: Holographic baryons and
instanton crystals, {\em Mod. Phys. Lett. B} {\bf 29} (2015) 1540052

\bibitem{KobNo} S. Kobayashi, K. Nomizu: Foundations of Differential Geometry, John Wiley \& Sons,
Inc. (Interscience Division), New York, Volume I,1963; Volume II,1969
 
\bibitem{KoS20} N. Kovensky, A. Schmitt: Holographic quarkyonic matter, {\em JHEP}
{\bf 09} (2020) 112

\bibitem{Kru90}  D. Krupka: Variational Sequences on Finite Order Jet Spaces, {\em Proc. Diff. Geom. Appl.}; J. Jany\v{s}ka, D. Krupka eds., World Sci. (Singapore, 1990) 236--254.

\bibitem{Kru15}  D. Krupka: Introduction to global variational geometry, {\em Atlantis Studies in Variational Geometry} {\bf 1} Atlantis Press 2015

\bibitem{ManSut04} N. Manton, P. Sutcliffe: Topological Solitons {\em Cambridge Monographs on Mathematical Physics}  Cambridge University Press, Cambridge (UK) 2004

\bibitem{MaMa92} K. B. Marathe, G. Martucci: The Mathematical Foundations of Gauge Theories; {\em Studies in Mathematical Physics} {\bf 5} North Holland, Amsterdam 1992

\bibitem{Noe18} E. Noether: Invariante Variationsprobleme, {\em Nachr. Ges. Wiss. G\"ott., Math. Phys. Kl.} {\bf II} (1918) 235--257.

\bibitem{NST16} A. Nakamula, S. Sasaki, K. Takesue: ADHM construction of (anti-)self-dual instantons in eight dimensions, {\em Nuclear Physics B} {\bf 910} (2016) 199--224.

\bibitem{Naka17} A. Nakamula: Atiyah-Manton construction of Skyrmions in eight dimensions, {\em J. High Energ. Phys} {\bf 2017} (2017) 076

\bibitem{ABC99} V. A. Novikov, M. A. Shifman, A. I. Vainshtein V. I. Zakharov: ABC of Instantons; in M. A. Shifman: ITEP Lectures on Particle Physics and Field Theory, Volume 1 {\em World Scientific Lecture Notes in Physics}, World Scientific, Singapore (1999) 201--299

\bibitem{PaRoWiMu16} M. Palese, O. Rossi, E. Winterroth, J. Musilov\'a: Variational sequences, representation sequences and applications in physics, {\em SIGMA} {\bf 12 } (2016) 045, 45 pages  (2016).

\bibitem{PaWi16} M. Palese, E. Winterroth: Topological obstructions in Lagrangian field theories, with an application to $3$D Chern--Simons gauge theory, {\em J. Math. Phys} {\bf 58 (2)} (2017) 023502 (2017).

\bibitem{PaWi22} M. Palese, E. Winterroth: A cohomological obstruction in higher dimensional Chern--Simons gauge theories {\em Int. Journ. Geom. Meth. Mod. Phys}  {\bf 19} (3) (2022) 2250032.

\bibitem{PoWu09} A. Pomarol and A. Wulzer: ``Baryon Physics in Holographic QCD'' {\em Nucl. Phys. B} {\bf 809} (2009)
347 

\bibitem{PaN7} B. Povh, K. Rith, C. Scholz, F. Zetsche, W. Rodejohann: Particles and Nuclei 7th ed. {\em Graduate Texts in Physics} Springer-Verlag Berlin Heidelberg 2015

\bibitem{RSZ10} M. Rho, S.-J. Sin, and I. Zahed: Dense QCD: a holographic dyonic salt,
{\em Phys.Lett. B} {\bf 689} (2010) 23,

\bibitem{SaSu1} T. Sakai and S. Sugimoto: Low energy hadron physics in holographic QCD {\em Prog. Theor.Phys.} {\bf 113} (2005) 843

\bibitem{SaSu2} T. Sakai and S. Sugimoto: More on a holographic dual of QCD, {\em Prog. Theor.
Phys.} {\bf 114} (2005) 1083

\bibitem{Saund} D. J. Saunders: The geometry of jet bundles, {\em London Mathematical Society Lecture Note Series} {\bf 142}, 
 Cambridge University Press, Cambridge (UK) 1989

\bibitem{Sta88} J. D. Stasheff: Cohomological Physics; in Y. Felix Ed.: {\em Algebraic Topology Rational Homotopy} {\bf LNM 1318} Springer-Verlag,  Berlin Heidelberg, 1988

\bibitem{Sut15} P. Sutcliffe: 'Holographic Skyrmions {\em Modern Physics Letters B} {\bf 29} (16) (2015) 1540051

\bibitem{Take17} K. Takesue: ADHM construction of (anti-)self-dual instantons in 4n dimensions, {\em J. High Energ. Phys.} {\bf 2017} (2017) 110

\bibitem{Tian00} G. Tian: Gauge theory and calibrated geometry I, {\em Ann. of Math.} {\bf (2) 151} (2000) 193--268.
 
\bibitem{Wit98} E. Witten: Anti de Sitter space and Holography {\em Adv. Theor. Math. Phys.} {\bf 2} (1998) 253-291

\bibitem{Zan12} J. Zanelli: Chern-Simons forms in gravitation theories {\em Classical and Quantum Gravity} {\bf 29} 133001 (2012)

\bibitem{Zum84} B. Zumino: Chiral Anomalies and differential Geometry; in B. S. Eds.: {\em Relativity, Groups
and Topology II, Les Houches 1983} North Holland, Amsterdam 1984

\end{thebibliography}
\end{document}